\documentclass[12pt]{article}

\usepackage{fullpage,amssymb,amsfonts,amsmath,xcolor,natbib,mathtools,longtable,pdflscape}
\usepackage{comment,bold-extra,pdfpages,soul,url}
\let\hat=\widehat
\def\diag{\text{diag}}
\def\pconv{\smash{\mathop{\longrightarrow}\limits^p}}
\def\dconv{\smash{\mathop{\longrightarrow}\limits^d}}
\def\dsim{\smash{\mathop{\sim}\limits^d}}
\def\var{\text{var}}

\newtheorem{lemma}{Lemma}
\newtheorem{proposition}{Proposition}

\def\yeit{Y_{it}(1)}
\def\ycit{Y_{it}(0)}

\def\var{\text{var}}
\def\cov{\text{cov}}
\def\blup{\textsc{\small blup}}

\def\plup{\textsc{\small plup}}
\def\pup{\textsc{\small pup}}
\def\plupi{\textsc{\small plupi}}
\def\plupd{\textsc{\small plupd}}

\begin{document}

\author{S\'{\i}lvia Gon\c{c}alves\thanks{%
McGill University, Department of Economics, 855 Sherbooke St. W., Montreal,
PQ, H3A,2T7. Email: silvia.goncalves@mcgill.ca} \and Serena Ng\thanks{%
Columbia University and NBER. Department of Economics, 420 W. 118 St. MC
3308, New York, NY 10027. Email: serena.ng@columbia.edu \newline
 We thank Atsushi Inoue, Simon Lee,  Jushan Bai, Kaspar W\"{u}thrich, Marcelo Medeiros, Bruno Ferman, Yuya Sasaki,  Bruce Hansen, Jeff Wooldridge, Pedro Sant'Anna, Barbara Rossi, and participants at the JBES session at the 2024 ASSA meeting and the Gary Chamberlain Seminar in Econometrics  for helpful comments. This work was supported by an NSERC grant and the National Science
Foundation (Ng SES 018369).}}
\title{Imputation of Counterfactual Outcomes when the Errors are Predictable}
\date{\today}
\maketitle

\begin{abstract}
A crucial input into causal inference is the imputed  counterfactual outcome.
 Imputation error can arise because of  sampling uncertainty from estimating the prediction model using the untreated observations, or from out-of-sample information not captured by the model. While the literature has focused on sampling uncertainty, it vanishes with the sample size. Often overlooked is the possibility that  the  out-of-sample error  can be informative about the missing counterfactual outcome if it is mutually or serially correlated. Motivated by the   best linear unbiased predictor (\blup) of \citet{goldberger:62} in a time series setting,  we  propose an improved predictor of potential outcome when the errors are correlated.  The  proposed \pup\;  is  practical as it is  not restricted to linear models,
  can be used with   consistent estimators already  developed, and  improves  mean-squared  error for a large class of strong mixing error processes. Ignoring predictability in the errors can  distort  conditional inference. However,  the precise impact  will depend on the choice of estimator as well as the realized values of the residuals.
%Controlling for  predictability  is particularly important in inference  of individual treatment effects when the out-of-sample prediction error cannot be averaged out.

\end{abstract}

\blup.


%Converges
%\newtheorem{proposition}{Proposition}[section]
\thispagestyle{empty} \setcounter{page}{0} \newpage \baselineskip=18.0pt
\setcounter{table}{0}
\section{Introduction}

Understanding the  effects of policies is an important aspect of economic analysis and many questions of interest involve an individual's or a group's response  to multiperiod interventions.  Given treatment status, a researcher observes  the outcome of unit $i$ after intervention at $t=T_0+1$ (denoted $Y_{it}(1)$) and wants to compare it to  the hypothetical outcome without intervention (denoted $\ycit$). Since we do not observe $\ycit$ for $t>T_0$,    these values need to be imputed.   Now imputation is concerned with the prediction of values  that will  never be  sampled, and from results in the  prediction literature,  we know that in-sample estimation uncertainty should diminish with the sample size and  what dominates  total prediction error asymptotically  is   the variation  not explained by the model. Yet,  in applications,  we tend to perform robust inference taking the residuals as given, when an improved  prediction is possible   by removing predictable variations that might still be in the residuals.

\begin{figure}[ht]
\hspace*{-1.0in}
\caption{Effect of German Unification,  GDP growth}
\label{fig:germany-diff}
\begin{center}
\includegraphics[width=7.0in,height=6.0in]{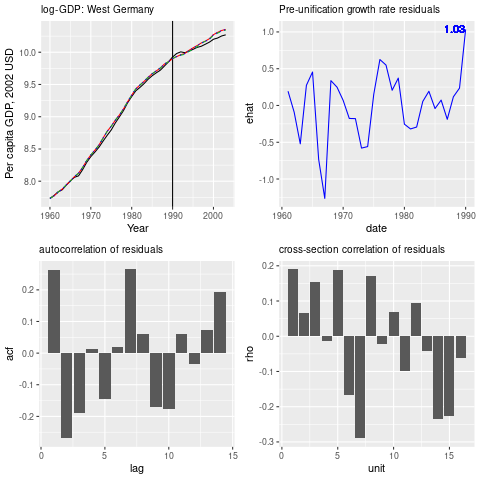}
\end{center}
The data are downloaded from \url{https://doi.org/10.7910/DVN/24714}

\end{figure}

\begin{figure}
\hspace*{-1.0in}
\caption{Effect of German Unification on level of GDP}
\label{fig:germany-level}
\begin{center}
\includegraphics[width=7.0in,height=6.0in]{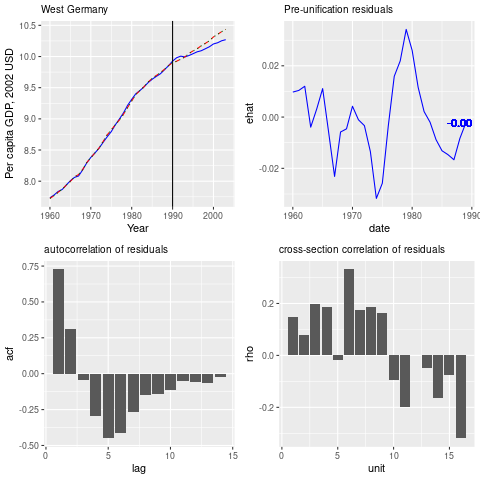}
\end{center}
\end{figure}

 To illustrate, consider Figure \ref{fig:germany-diff}  which studies the  impact of the  German reunification in 1990 on $Y_{1t}=\log$ GDP. Because the GDP data are  non-stationary, we  estimate common factors from a 16 country panel  of  GDP growth  $(\Delta Y_{2,1:T}(0),\ldots, \Delta Y_{17,1:T}(0))$,  where $\Delta Y_{i,1:T}(0)\equiv (\Delta Y_{i1}(0),\ldots, \Delta Y_{iT}(0))'$.  The top-left panel  displays  actual (log) GDP  over the full sample, along with the in-sample fit $\hat Y_{1,1:T_0}(0)$, and the   counterfactual values $\hat Y_{1,T_0+1:T}(0)$.   The effect of  reunification on  GDP is stark, but masks  the fact that the (in-sample)  residuals  $\hat e_{1,1:T_0}=\Delta Y_{1,1:T_0}(0)-\hat{\Delta Y}_{1,1:T_0}(0)$ are  persistent. This can be  seen from the correlogram in the bottom left panel, or from   the plot of the series itself  in the top right panel.  The series $\hat e_{1,1:T_0}$ is also cross-correlated with other errors, though many of the  $\text{corr}(\hat e_{1,1:T_0},\hat e_{j,1:T_0})$ are not statistically significant as shown  in the bottom right panel.   The in-sample residuals of the log level model are also serially correlated, as   shown in Figure \ref{fig:germany-level}.   Time series and  cross-section  correlation of  the in-sample residuals  is not specific to this example.
%\footnote{We also analyzed  the tourism data for the Basque country prior  to the outbreak of terrorism,   and the cigarette sales in California prior to Proposition 99, using different imputation methods.  Though residual  variance relative to the trending series tends to be small, the residuals are persistent.}

This paper considers the implications of non-spherical errors for model-based imputation. Non-spherical errors, which induce predictability,  can arise because the model for $\ycit$ is mis-specified or because  $\ycit$ cannot be adequately captured by observed information without further signal extraction.  We build on   the  best-linear-unbiased predictor (hereafter, \blup) developed in \cite{goldberger:62}
for  linear models with non-spherical errors.  The key to \blup\; is not that it is based on GLS estimation, but that it  has a correction term that depends on the covariance  structure of the errors.
We suggest  a  practical predictor (\plup) that is asymptotically equivalent to the infeasible \blup\; to a first order. Furthermore, if predictability is due to serial correlation, a simple AR(1) correction will  reduce the mean-squared prediction error for a large class of stationary mixing error processes. This is not to say that an AR(1) correction is best as  the desired adjustment  will  necessarily be data dependent, but to   point out  that simple modifications can reduce the mean-squared error of  the  standard  prediction.

We adapt Goldberger's result for  linear prediction  to an imputation setting when  the  counterfactual outcomes are never  observed.  We derive  infeasible \blup\; for  linear panel data models and make precise its dependence on the covariance structure of the errors. We  show that when $T_0$ is large,  a \plup\; that controls for time and/or  cross-section correlation can be constructed.  But the idea of correcting the standard prediction for predictable errors is more general and can be applied to  non-linear models when direct modeling of dynamics   is not so straightforward. Thus when linearity is not required,  we   refer to the practical unbiased predictor as \pup.\footnote{We thank Bruce Hansen for this suggestion.}

In addition to inefficient point predictions, ignoring correlation in the errors  also has implications for  inference. Though a standard prediction will yield asymptotically  valid  unconditional inference, the prediction interval will be wider because correlated residuals inflate  error variance. More concerning is that standard prediction  is  biased conditional on pre-treatment outcomes of both the treated and the untreated, as well as the post-treatment outcomes of the untreated. This bias may  distort inference and   the precise impact will depend not just on  persistence of the residuals, but also on   the realized values of the residuals relevant for imputing $Y_{i,T_0+h}(0)$.
In the German unification example,  $\hat e_{1,T_0}$ is  1.03 in the growth rate model. This non-zero value yields  \pup\;  growth rates that are  slightly different from  the standard prediction.
% as shown in   Figure \ref{fig:germany-growth}.

\pup\;  is concerned with reducing the out-of-sample error and  does not preclude the use of  robust  standard errors  or    resampling schemes to account for correlation in the in-sample residuals. In practice, a  \pup\;  for unit $i$ will use residuals  of  unit $i$  before treatment, and possibly  of the untreated units after $T_0$. For  serial correlation type error dependence that should die off as $h$ increases, a  \pup\; correction is most effective in  imputating $Y_{i,T_0+h}(0)$ at small $h$. Recent work by  \citet{cwz-jasa:21}, \citet{fmm-jasa:22}, and \citet{ferman-jae:23} can be seen from  a  \pup\; perspective.
% using pre-treatment data,  such as discussed in  \citet{bdm-qje:04},  \citet{hansen:07}, \citet{wooldridge-twfe:21},  and \citet{borusyak:22}.

Our central message is that in-sample uncertainty is asymptotically dominated by variability of the out-of-sample prediction error, and more attention should be paid  to  improving the point-prediction before turning to inference.  The rest of the paper proceeds as follows. Section 2 sets up the econometric framework and provides motivating examples for predictable errors. Section 3  summarizes the
properties of \blup\; and then presents  \plup. Its   mean-squared error is analyzed using  asymptotic expansions of the population prediction error under mixing  conditions. Section 4 switches focus from predicting future outcomes   to imputing missing values.
 Unconditional and conditional coverage of  prediction intervals  are analyzed in Section 5.    With some abuse of language, we sometimes use   'imputation' and 'prediction' interchangeably. Our discussion will  focus on  time dependence but the arguments also hold for spatial and cross-section dependence.

\section{The Econometric Framework}
We will use the standard potential outcome framework for analysis.
  Let    $\yeit$   be the potential response for unit $i$ at time $t$ if it was  exposed to treatment (or policy intervention), and  $\ycit$ be the potential response of a (control) unit $i$ that was  not exposed to intervention at $t$.   We observe $Y_{it}=\ycit(1-D_{it})+\yeit D_{it}$ where treatment status  $D_{it}=1$  if unit $i$ is exposed in period $t$  and is zero otherwise.
 Without loss of generality, we order  the $N_1$     exposed units  before the  $N_0=N-N_1$ unexposed units.   We  observe
\[ Y_{it}=\begin{cases}
\ycit, \quad    &i=1,\ldots,  N,\; \quad\quad\quad t=1,\ldots, T_0\\
\ycit, \quad    &i=N_1+1,\ldots, N, \;\;\;  t= T_0+1,\ldots, T\\
 \yeit  & i=1,\ldots, N_1, \; \quad\quad\;\; t=T_0+1,\ldots, T
\end{cases}
\]
 and are interested in  the   effect on unit $i\in[1,N_1]$  in  $h>0$ periods after treatment at $T_0$. Different   average  treatment effects  can be derived
 from the individual  treatment effect,  defined as
\[ \delta_{i,T_0+h}=\underbrace{Y_{i,T_0+h}(1)}_{\substack{\text{observed outcome }\\ \text{h periods after  treatment }}}-\underbrace{Y_{i,T_0+h}(0)}_{\substack{\text{ outcome without treatment}\\ \text{in period } T_0+h}}.\]
The econometrics challenge is that $Y_{it}(0)$ is not observed for $i\le N_1$ when $t>T_0$.

Following the literature, we assume that $\ycit$ has a pseudo-true conditional mean (or mean-unbiased proxy) $m_{it}=\mathcal M(\beta; \mathcal H)$ that is parameterized by a vector $\beta$ given some information set $\mathcal{H}$, and $e_{it}=Y_{it}(0)-m_{it}$ is such that $E(e_{it})=0$. For example, an AR(1) approximation would make $m_{it}=\rho y_{it-1}$ and $\mathcal H$ would be $y_{is}$ for  $s\le T_0$. Being a pseudo-true mean,  $m_{it}$  may not coincide with the true conditional mean say,  $m^*_{it}$, where  $e^*_{it}=Y_{it}(0)-m^*_{it}$.
 For each $i=1,\ldots, N_1$,
\begin{eqnarray*}
\ycit&=& m_{it}+e_{it}, \quad\quad\quad\quad t=1,\ldots, T\\
\yeit&=& m_{it}+\delta_{it}+e_{it},\quad\quad t>T_0.
\end{eqnarray*}
Let  $\hat m_{it}$ be a consistent estimate of $m_{it}$. Then
\[ \ycit=\hat m_{it}+\hat e_{it}. \]
Since $\hat Y_{it}(0)=\hat m_{it}$,  the  treatment  effect on unit $i$ at a given $t=T_0+h$ is then estimated by \begin{eqnarray*}
 \hat \delta_{i,T_0+h}&=& Y_{i,T_0+h}(1)-\hat Y_{i,T_0+h}(0)\\
&=& Y_{i,T_0+h}(1)-Y_{i,T_0+h}(0)+ (Y_{i,T_0+h}(0)-\hat m_{i,T_0+h})\\
&=& \delta_{i,T_0+h}+e_{i,T_0+h}+ (m_{i,T_0+h}-\hat m_{i,T_0+h}).
\end{eqnarray*}
  The pointwise imputation/prediction  error is
\begin{eqnarray*}
\hat \delta_{i,T_0+h}-\delta_{i,T_0+h} &=& (m_{i,T_0+h}-\hat m_{i,T_0+h})+e_{i,T_0+h}.
%&=& \hat e_{1,T_0+h}
\end{eqnarray*}
This error has two sources of variation: one  from in-sample estimation of  $m_{it}$, and one due to the  out-of-sample error $e_{i,T_0+h}$ not captured by the model.
The first error will be  negligible as $T_0$ increases provided that $\hat m_{it}$ is consistent for $m_{it}$ in some well defined sense, but  the second error does not vanish with the sample size and thus total prediction error variance is minimized asymptotically if $m_{it}$ is chosen  such $e_{it}$ does not contain predictable information. However,
 theory actually allows $e_{it}$ to be serially and/or mutually correlated, and while the assumption of no correlation is convenient, it is not always appropriate. In the next subsection, we provide some examples for dependence in the errors.

\subsection{Examples when $e_{it}$ is predictable}
We will first clarify what we mean by in-sample and out-of-sample errors. To fix ideas, suppose that unit 1 is being treated  and  the model is linear so that $m_{1t}=x_{t}'\beta$.  Single equation estimation yields the imputed value $\hat \delta_{1,T_0+1}=x_{T_0+1}'\hat\beta$ and imputation error
 $ \hat \delta_{1,T_0+1}-\delta_{1,T_0+1}= -x_{T_0+1}'(\hat\beta-\beta) +e_{1,T_0+1}$ whose  variance is
 \[\text{var}(\hat \delta_{1,T_0+1}-\delta_{1,T_0+1})=\sigma_e^2+ x_{T_0+1}' \text{var}(\hat\beta) x_{T_0+1}.\]  Correlation in $x_te_t$ may  necessitate  robust standard errors for $\hat\beta$, but provided that $E[x_te_{1t}]=0$, $\hat\beta$ is consistent in the sense that   $\var(\hat\beta)\rightarrow 0$ as $T_0\rightarrow \infty$. Thus, the variance of  imputation error
  is  dominated by the  out-of-sample error variance  $\sigma^2_e\equiv\var(e_{1,T_0+1})$ asymptotically. This variance is minimized when  $e_{1,T_0+1}$ is uncorrelated. Serial correlation  can arise because of temporal aggregation, but residual correlation (temporally or mutually) is usually a symptom of  misspecification of the model or conditioning information.  We give some examples below.

\paragraph{Example misspecification  1:} Suppose that $Y_{1,t}(0)=\phi_1 Y_{1,t-1}(0)+\phi_2 Y_{1,t-2}(0)+v_{1t}$ is an AR(2) process with iid innovations $v_{1t}$, but the researcher assumes an AR(1) model. Then $m_{1t}= \beta Y_{1,t-1}(0)$,  the  pseudo true parameter is $\beta=\frac{\phi_1}{1-\phi_2}\ne \phi_1$, and
 $e_{1t}=v_{1t}+(\phi_1-\beta) Y_{1,t-1}(0)+\phi_2 Y_{1,t-2}(0)$  is serially correlated  when $\phi_2\ne 0$.

\paragraph{Example  misspecification 2:} Suppose that the potential outcome has an  interactive  fixed effect structure: $Y_{1t}(0)=\lambda_1'F_t+\epsilon_{1t}$ but a  researcher specifies an additive fixed effect model $Y_{1t}(0)=\lambda_1+F_t+e_{1t}$. Then  $e_{1t}$ will be serially correlated if $F_t$ is serially correlated,  even if $\epsilon_{1t}$ is white noise.

\medskip

 Factor-based imputation  assumes  $X_{it}=\lambda_i'F_t+e_{it}$ where  $F_t$ is a vector of $r$ latent common factors with $\lambda_i$ as loadings and $e_{it}$ is an idiosyncratic error.
An appeal of factor-based imputation is that under some conditions,  the space spanned by $F$ can be consistently estimated without modeling  the (weak cross-section or time) dependence in the idiosyncratic errors.
While  \citet{xu:17}  iteratively estimates $F$ and the missing values jointly by  principal components (PCA),  \citet{baing-jasa:21} impute the missing values (or complete the  matrix) using two full-sample applications of PCA.      \citet{athey-jasa:21, syndid-aer:21} estimate the low rank component using singular value thresholding (SVT).\footnote{Regularization is not necessary to consistently estimate the missing values, but could give a lower rank  common component  than the one in \citet{baing-jasa:21}.} Though all consistent estimators of $F$ imply that
 $\hat F_{T_0+h}$ can be used as though they were observed predictors, the possibility remains that  $e_{1,T_0+h}$ can be predicted by information available.

%\paragraph{Example Omitted Information 1:}  Suppose that $Y_{1t}(0)=\Lambda_iF_t+  X_{1t}\theta+\epsilon_{it}$ where  $X_{1t}$ is a serially correlated regressor orthogonal to $F_t$ and $\epsilon_{1t}$.  The researcher assumes  $m_{1t}=\Lambda_1'F_t$ and   $e_{1t}=\epsilon_{1t}+ X_{1t}\theta $ is predictable by $X_{1t}$.

\paragraph{Example Correlated Idiosyncratic Errors} (from \citet{fmm-jasa:22}) Suppose that $Y_{it}(0)=\lambda_i'F_t+e_{it}$ and the researcher correctly assumes $m_{1t}=\lambda_1'F_t$ but $e_{it}$ is correlated with $e_{jt}$ for  $j$ in some index set $C$. Then  $e_{1t}=v_{1t}+\sum_{j\in C} \theta_j e_{jt}$ is predictable by those $e_{jt}$ where $j\in C$.

\bigskip

The method of
synthetic control (SC)  developed in \citet{abadie:03}  assumes that  there exist weights $\beta_j^*$  such that  a perfect fit $Y_{1t}(0)=\sum_{j=2}^{N} \beta^*_j Y_{j,t}(0)$ exists  for every $t\le T_0$.  \citet{abadie-jasa:10} make additional use of   $K$ economic predictors $X_t=(X_{1t},X_{0t})$, where  $X_{0t}=(X_{2t},\ldots, X_{N,t})$ for the unexposed.    The Synthetic Difference-in-Difference (SDID) in \citet{syndid-aer:21} also  reweights the  pre-treatment time periods  to balance the pre-and post exposure time periods and nests SC and  DID as special cases. However,
an increasing number of papers suggest that an  `imperfect pretreatment fit' may prevent recovery of  $\beta^*$.

\paragraph{Example  Imperfect Fit 1:} (from \citet{bfr:21})
 Suppose that  $Y_{it}(0)=\phi_1 Y_{i,t-1}(0)+v_{it}$ for all $i=1,2,\ldots,N$, and one constructs $m_{1t}=\sum_{j=2}^{N}\beta^*_j Y_{j,t}$. We can show that the error $e_{1t}=Y_{1t}(0)-m_{1t}$ can be decomposed as
%=\phi_0+\phi_1 Y_{1,t-1}(0)+v_{1,t}- \sum_{j=2}^{N+1} \beta^*_j Y_{j,t}$
$e_{1t}
=\phi_1 e_{1,t-1} +v_{1t}-\sum_{j=2}^{N}\beta^*_j v_{jt}$. The error $e_{1,T_{0}+1}$ contains an imbalance component $\phi_1 e_{1,T_0}=\phi_1 (Y_{1,T_0}-\sum_{j=2}^N \beta^*_j   Y_{j,T_0})$ (which is zero if there is perfect fit but not otherwise)
as well as a noise component $v_{1,T_0+1}-\sum_{j=2}^{N} \beta^*_j v_{j,T_0+1}$, and both can contribute to serial correlation.

%\medskip

\paragraph{Example Imperfect Fit 2:}  (from \citet{ferman-pinto:21}) Suppose that
$Y_{it}(0)=c_i+\Lambda_i'F_t+\epsilon_{it}$ and one estimates $\hat\beta=\text{argmin}_{b}  \|Y_1(0)-X_0b)||_2^2$ where $X_0=(Y_2,\ldots, Y_N)$. With  $\beta^*=\text{plim} \hat\beta$, the population imputation error  is
\[ e_{1t}=Y_{1t}(0)-X_{0t}'\beta^*=\bigg(c_1-\sum_{j=2}^{N}\beta^*_j c_j \bigg)+F_t'\bigg(\lambda_1-\sum_{j=2}^{N} \beta_j^* \lambda_j\bigg)+\bigg(\epsilon_{1t}-\sum_{j=2}^{N} \beta^*_j \epsilon_{jt}\bigg).\]
The first two terms  vanish only if $\sigma_{\epsilon}\equiv \var(\epsilon_{it})= 0$; otherwise,   $(c_1,\lambda_1)\ne (\sum_{j=2}^N \beta_j^*c_j, \sum_{j=2}^N \beta_j^*\lambda_j)$.
\citet{ferman-pinto:21} suggest to remove the bias with a  mean adjustment but this may not remove serial or mutual correlation in the errors.

In the above examples,  $e_{it}$   absorbs all sorts of deficiencies in $m_{it}$  and  thus  contains information about $Y_{it}(0)$.   Cross-section, spatial, and  time dependence in $e_{it}$ are examples of non-spherical errors.

\section{Prediction with Non-Spherical Errors}

This section uses classical results in linear prediction to motivate how information in the errors can be used to improve prediction.
We will consider  optimal linear prediction  in a  time series  setting so that  the $i$ subscript can be dropped.

\subsection{Goldberger's \small{BLUP}}
This subsection  summarizes results for  best linear unbiased prediction, \blup.
The concept  seems to be first considered in \citet{henderson:50}  in the animal breeding literature to predict the quality of offsprings. It is still widely used in  estimation of random effects in linear mixed models for cross-section data.\footnote{\citet{robinson:91} provides a survey of its many derivations, including a Kalman filter interpretation, see also \citet{spall:91}.  \citet{taub:79} and \citet{baltagi-jf:08,baltagi-handbook} use it in variance components analysis of  panel data.}  \citet{goldberger:62}   formalizes the idea in a setting where  the linear model for predicting  a  scalar variable $y_t$ is given by
\begin{equation}  \label{dgp}
y_t=X_t^{\prime }\beta+e_t
\end{equation}
where  $X_t$ is a $K\times 1 $ vector of completely
observed predictors assumed to be fixed in repeated samples, $\beta$ is a vector of time invariant parameters,
$e_{t}$ is a zero mean stationary process that is possibly serially correlated,  and $ \Omega$ is the $T_0\times T_0$ covariance matrix of the $T_0\times 1$ vector $e$.

\citet{goldberger:62} is
interested in a linear unbiased prediction of $y_m$  at some  $m>T_0$  given information up to $T_0$  when $\Omega$ is positive definite but has non-zero off-diagonal entries. Let $X$ be at $T_0\times K$ matrix of regressors. The assumption of
squared loss  $E[(y_{m}-y_{m|T_0})^{2}]$ implies a linear predictor of the form $%
y_{m|T_0}=A^{\prime }y$ with prediction error $y_{m|T_0}-y_{m}=(A^{%
\prime }X-X_{m}^{\prime })\beta +A^{\prime }e-e_{m}$. The unbiasedness constraint  $E[y_{m|T_0}]=y_{m}$ requires that  $%
A^{\prime }X-X_{m}^{\prime }=0$, implying  a prediction variance of
\begin{eqnarray*}
\sigma _{m}^{2} &=&E[(y_{m|T_0}-y_{m})^{2}]=E[A^{\prime }ee^{\prime
}A+e_{m}^{2}-2A^{\prime }e_{m}e] \\
&=&A^{\prime }\Omega A+E\left( e_{m}^{2}\right) -2A^{\prime }\omega
\end{eqnarray*}%
where $\omega \equiv E[e_{m}e]$. Let $\lambda$ be the Lagrange multiplier on the unbiasedness constraint.  Minimizing $A^{\prime }\Omega A-2A^{\prime
}\omega -2\lambda ^{\prime }(X^{\prime }A-X_{m})$ with respect to $A$
%with FOC
%\begin{eqnarray*}
%0&=&2\Omega A-2\omega-2X\lambda\\
%0&=&2X'A-2x_m
%\end{eqnarray*}
gives the best linear unbiased prediction (\blup)
\begin{equation*}
y_{m|T_0}^{\ast }=x_{m}^{\prime }\beta _{GLS}+\omega ^{\prime }\Omega ^{-1}{e}_{GLS},
\end{equation*}%
where $\beta _{GLS}=(X^{\prime }\Omega ^{-1}X)^{-1}X^{\prime }\Omega ^{-1}y$
is the infeasible GLS estimator, and ${e}_{GLS}=y-X\beta _{GLS}$ is a $T_0\times 1$
vector of  errors.  Notably,    $%
y^*_{m|T_0}$ depends on assumptions about $\Omega $, and in a time series setting, this depends on the dynamics  of $e_t$. If $e_{t}=\phi_{1}
e_{t-1}+v_{t}$, where $v_t\dsim (0,\sigma^2_v)$,  $\Omega$ is  $\sigma^{2}_{e}=\sigma^2_{v}/(1-\phi_{1}^2)$  times  a $T_0\times T_0$  Toeplitz matrix with $\phi_{1}^{i-1}$ on the $i$-th diagonal.
%\begin{equation*}
%\Omega =\sigma ^{2}_{e}\left(
%\begin{array}{cccc}
%1 & \phi_{1} & \ldots & \phi_{1} ^{T-1} \\
%\vdots & 1 &  & \phi_{1} ^{T-2} \\
%\phi_{1} ^{T-1} & \ldots &  & 1%
%\end{array}%
%\right) ,
%\end{equation*}%
Then $\omega \equiv E[e_{m}e]=\phi_{1} ^{-T_0+m}\Omega _{T_0}$, where $\Omega _{T_0}$ is
the last column of $\Omega $. The AR(1) assumption implies a \blup\;at $m=T_0+h$ of
\begin{equation*}
y_{T_0+h|T_0}^{\ast }=X_{T_0+h}^{\prime }\beta _{GLS}+\phi_{1} ^{h}{e}_{GLS,T_0}
\end{equation*}
with prediction error  $e^*_{T_0+h|T_0}=y_{T_0+h}-y_{T_0+h|T_0}^{\ast
}=e_{T_0+h}-\phi^h_{1} e_{T_0} +o_p(1)=
v_{T_0+h}+o_p(1)$ where the $o_p(1)$ term converges to 0 as $T_0\rightarrow\infty$.
%\footnote{For example, in the one-way error component case when $w_{it}=w_i\sim (0,\sigma^2_w)$,  $\Gamma= \sigma^2_w (I_N\times J_{T_0})+\sigma^2_v(I_N\times I_{T_0})$, where $J_{T_0}$ is a matrix of ones of dimension $T_0$, while $I_N$ and $ I_{T_0}$ are identity matrices of dimension $N$ and $T_0$, respectively.}

 \blup\; is infeasible because $\phi_{1} $ is not observed.  Feasible \blup\; requires  iterative Cochrane-Orcutt or Prais-Winsten estimation of $\phi_{1}$, or  direct estimation of $\beta$ and $\phi_{1}$ from a  Durbin equation.\footnote{Cochrane-Orcutt performs  least squares
regression of $y_{t}-\phi_{1} y_{t-1}$ on $X_{t}-\phi_{1} X_{t-1}$ for given
$\phi_{1} $ using data from $t=2,\ldots ,T_0$, and then estimates $\phi_{1} $
from an autoregression in $y_{t}-X_{t}^{\prime }\hat{\beta}$ till
convergence. The Prais-Winsten estimator additionally exploits information
in $t=1$. It is also possible to estimate $\beta$ and $\phi_{1}$ directly
from the Durbin equation $y_{t}=X_{t}^{\prime }\beta+ y_{t-1}\phi_{1}+X_{t-1}^{\prime }{\gamma}+$%
error.}  These feasible estimators are all efficient and consistent.  \citet{cochrane-orcutt:49}  suggest to improve the standard prediction by incorporating lags of regressors and the dependent variable. The difference is that $\hat e_{GLS,t}$  summarizes the dynamic relation between $y$ and $X$ into a single signal and can be appealing when $X$ is of high dimension.

A feasible \blup\;differs from the OLS prediction in two ways. First, it uses $\hat{\beta}%
_{GLS}$ instead of $\hat{\beta}_{OLS}$ and thus requires the dynamics of $e_t$ to be specified.  Second,
 \blup\; adds to the GLS prediction a term that  adjusts for serial correlation
in $e$ which in this AR(1) example is  $\phi_1 e_{T_0}$ for $h=1$. Since \blup\; is an optimal prediction, it is more efficient than an OLS prediction.  To make this point precise, consider again  the AR(1) case. At $h=1$, feasible \blup\;
\begin{eqnarray*}
\hat{y}_{T_0+1|T_0}^{\ast }&=&X_{T_0+1}^{\prime }\hat{\beta}_{GLS}+\hat{\phi_{1}}%
\hat{e}_{GLS,T_0}
\end{eqnarray*}%
has prediction error $\hat e_{T_0+1|T_0}^*=y_{T_0+1}-\hat y_{T_0+1|T_0}^*$, or
\begin{eqnarray}
\hat{e}_{T_0+1|T_0}^{\ast }&=&v_{T_0+1}-\left( X_{T_0+1}-\phi_{1} X_{T_0}\right)
^{\prime }(\hat{\beta}_{GLS}-\beta )-\left( \hat \phi_{1,GLS}-\phi_{1}
\right) \hat{e}_{GLS,T_0} \nonumber  \\
&=&v_{T_0+1}+o_{p}(1) \label{eq:blup-v}
\end{eqnarray}%
where the $o_{p}(1)$ term comes from the fact that the jointly estimated $%
\hat{\beta}_{GLS}$ and $\hat\phi_{1,GLS}$ are $\sqrt{T_0}$ consistent for $%
\beta $ and $\phi_{1} $. Since $\hat e_{T_0+1|T_0}^*$ is asymptotically $v_{T_0+1}$
whose variance is $\sigma^2_v$, feasible \blup\; achieves the same asymptotic
efficiency as infeasible \blup.

In contrast, the OLS prediction error is
\begin{eqnarray}
\hat{e}_{T_0+1|T_0}&=&e_{T_0+1}-X_{T_0+1}^{\prime }(\hat{\beta}_{OLS}-\beta
)\nonumber \\
&= &e_{T_0+1}+o_{p}(1)  \label{eq:ols-v}
\end{eqnarray}%
where the $o_p(1)$ term comes from $\sqrt{T_0}$ consistency of $\hat{\beta}_{OLS}$ for  $\beta$. But $\hat
e_{T_0+1|T_0}$ is asymptotically $e_{T_0+1}$ whose variance is $\sigma^2_e\ge
\sigma^2_v$. Thus, the MSE
improvement of \blup\; over OLS is  due to  the additional term $\phi_{1} e_{T,GLS}$ in the prediction,  not because of GLS versus OLS estimation per se. Building on this idea, we
will consider a  linear
prediction that is also asymptotically unbiased but can  improve upon the OLS prediction without a priori  knowledge of the precise dynamic structure of $e_t$.

%\subsubsection{A Practical Linear Unbiased Prediction (\plup)}

%Summary of notation: $  y^{\ast}_{T+1|T}=GLS,$ $\hat y_{T+1|T}$ is OLS, '+' means correction, $\tilde  y_{T+1|T}$ is any other consistent estimator of $\beta$.

\subsection{From \small{BLUP} to \small{PLUP} }
This subsection suggests a practical variant (\plup) and studies its mean-squared error (MSE) using asymptotic expansions, first for $h=1$, and then for  $h>1$ when direct and iterative forecasts are possible.

Our point of departure is that any predictor that controls for serial
correlation will have the same first order effect as feasible \blup. Let $%
\hat\beta$ denote the least squares estimate of $\beta$. Consider modifying
the (standard) least-squares prediction $\hat y_{T_0+1|T_0}=X_{T_0+1}^{\prime}\hat{%
\beta}$ as follows:
\begin{equation}
\hat{y}^+_{T_0+1|T_0}=\hat y_{T_0+1|T_0}+\hat\rho_{1}\hat{e}_{T_0}  \label{eq:plup}
\end{equation}%
where $\hat{e}_{T_0}=y_{T_0}-X_{T_0}^{\prime}\hat{\beta}$ is the OLS residual, and
\begin{equation}  \label{eq:rho_hat}
\hat {\rho_{1}}=\frac{\sum_{t=1}^{T_0}\hat {e}_{t-1}\hat {e}_{t}}{%
\sum_{t=1}^{T_0}\hat {e}_{t-1}^{2}} \\
\end{equation}
is the least squares estimate of the first order autocorrelation coefficient
of $\hat e_{t}$. Note that unlike feasible \blup\;which
re-estimates $\beta$ after $\hat\rho_{1}$ is available,  we simply adjust the OLS
prediction $\hat y_{T_0+1|T_0}$ for serial correlation with $\hat\rho_1\hat e_{T_0}$. The prediction error  $\hat e_{T_0+1|T_0}^+= y_{T_0+1}-\hat{y}^+_{T_0+1|T_0}
= y_{T_0+1}-\hat y_{T_0+1|T_0}-\hat\rho_{1}\hat e_{T_0}$ is
\begin{eqnarray*}
\hat e_{T_0+1|T_0}^+&=&\hat e_{T_0+1|T_0} -\hat\rho_{1}\hat e_{T_0} \\
&=& e_{T_0+1}-\rho_{1} e_{T_0}+o_{p}(1),
\end{eqnarray*}
where the last equality follows because $\hat\beta\pconv \beta$ and
$\hat \rho_{1}\pconv \rho_1$. If $e_t$ is indeed an AR(1) model, then $%
\rho_{1}=\phi_{1}$ and
\begin{eqnarray*}
\hat e_{T_0+1|T_0}^+ &=&v_{T_0+1}+o_{p}(1),
\end{eqnarray*}%
which is asymptotically equal to $v_{T_0+1}$, the prediction error of the
infeasible \blup.

As it turns out, adding the term $\hat{\rho}_{1}\hat{e}_{T_0}$ to any
consistent prediction $\hat{y}_{T_0+1|T_0}$ will  yield an efficiency gain
even when the true model is not an AR(1). We  will refer to the
prediction
\begin{equation*}
\hat{y}_{T_0+1|T_0}^{+}=X_{T_0+1}^{\prime }\hat{\beta}+\hat{\rho}_{1}\hat{e}_{T_0}
\end{equation*}%
as practical \blup\; (or \plup), practical because it does not require GLS estimation and it is asymptotically as
efficient as \blup.  To formalize the
properties of the \plup\;  error $\hat{e}_{T_0+1|T_0}^{+}=y_{T_0+1}-\hat{y}%
_{T_0+1|T_0}^{+}$, we assume the following.

\begin{description}
\item[Assumption A1]

\item[(a)]   $E\left\vert e_{t}\right\vert ^{r}<\infty $ for some $r>2$, for
all $t$.

\item[(b)] $\left\{ e_{t}\right\} $ is a  zero mean strictly stationary strong mixing
process with mixing coefficients $\alpha \left( k\right) =O\left( k^{-\frac{r%
}{r-2}-\delta }\right) $ for some $\delta >0$.
\end{description}
We define the strong mixing coefficients as $\alpha \left( k\right)
=\sup_{A,B}\left\vert P\left( A\cap B\right) -P\left( A\right) P\left(
B\right) \right\vert $ where $A$ and $B$ vary over events in the sigma
fields generated by $\left\{ e_{s}:s\leq 0\right\} $ and $\left\{
e_{s}:s\geq k\right\} $, respectively.  Assumption A1 includes  linear processes $%
e_{t}=\sum_{j=0}^{\infty }\psi _{j}v_{t-j}$, where $\sum_{j=0}^{\infty
}\left\vert \psi _{j}\right\vert <\infty $ and $v_{t}$ is i.i.d.$\left(
0,\sigma _{v}^{2}\right) $ with $E\left\vert v_{t}\right\vert ^{r}<\infty $,
which includes  stationary invertible ARMA(p,q)
processes
and nonlinear weakly dependent processes with GARCH and ARCH innovations.

\begin{lemma}
\label{lem:lemma1} ($h=1$): Let $X_t$ be predictors and $e_{t}$ be the errors in the model defined by (\ref%
{dgp}). Suppose that $\{e_{t}\}$ satisfies Assumption A1 and that for $j=0,1$%
, $E\left( X_{t}e_{t-j}\right) $, $E\left( X_{t-j}e_{t}\right) $ and $%
E\left( X_{t}X_{t-j}^{\prime }\right) $ exist such that  $\hat{\beta}\pconv \beta $. Then  as $T_0\rightarrow \infty$,

\begin{itemize}
\item[(i)] $\hat{\rho}_{1}\smash{\mathop{\longrightarrow}\limits^p}\rho
_{1}\equiv \frac{\gamma _{1}}{\gamma _{0}}$, where $\gamma _{k}\equiv
E\left( e_{t}e_{t-k}\right) $ for all $k$;

\item[(ii)] Standard prediction error:  $\hat{e}
_{T_0+1|T_0}=e_{T_0+1}+o_{p}(1)$ where $e_{T_0+1}\dsim (0,\gamma_0)$;
% $E[e_{T_0+1}]=0$,  $\var(e_{T_0+1})=\gamma _{0}$.

\item[(iii)]  \plup\;error:  $\hat{e}_{T_0+1|T_0}^{+}=e_{T_0+1}-\rho
_{1}e_{T_0}+o_{p}(1)$ where $e_{T_0+1}-\rho_1 e_{T_0}\dsim (0,\gamma_0(1-\rho_1^2))$.
% $E[e_{T_0+1}-\rho _{1}e_{T_0}]=0$ and $%
%\var(e_{T_0+1}-\rho _{1}e_{T_0})=\gamma _{0}(1-\rho _{1}^{2})\leq \gamma _{0}$.

\end{itemize}
\end{lemma}

Part (i) shows that $\hat{\rho}_{1}$ converges to the first order
autocorrelation coefficient of $e_{t}$. Parts (ii) and (iii) describe the
asymptotic expansion of the prediction errors ignoring the estimation error
uncertainty. Part (ii) implies that the standard prediction is asymptotically
unconditionally unbiased in spite of not accounting for serial correlation because $
e_{T_0+1}$ is mean zero by assumption,  and it has  asymptotic
variance  $\gamma _{0}\equiv \text{var}\left( e_{T_0+1}\right) =\sigma
_{e}^{2}$. The \plup\; error in (iii) also has an unconditional mean of zero, but its
variance is $\gamma _{0}(1-\rho _{1}^{2})$. Since $\left\vert \rho
_{1}\right\vert \leq 1$,
\begin{equation*}
\gamma _{0}(1-\rho _{1}^{2})\le\gamma _{0},
\end{equation*}%
implying that the \plup\;  mean-squared dominates the standard
prediction.   If $e_{t}$ is truly generated as $e_{t}=\phi _{1}e_{t-1}+v_{t}$,
 the \plup\;error $\hat{e}_{T_0+1|T_0}^{+}$ will  be  asymptotically serially uncorrelated since $\rho _{1}=\phi_1$.

However, an AR(1) correction will improve upon   the standard prediction even  when $e_t$ is not an AR(1),  provided that $e_t$ is a mixing process satisfying A1.  For
instance, if $e_{t}$ is an AR(2) defined by $e_{t}=\phi _{1}e_{t-1}+\phi
_{2}e_{t-2}+v_{t}$, then $\rho _{1}=\frac{\phi _{1}}{1-\phi _{2}}\neq \phi
_{1}$. But it will still be the case that $\hat{\rho}_{1}\smash{\mathop{%
\longrightarrow}\limits^p}\rho _{1}$ as stated in (i).  The \plup\;
error is now $\hat{e}_{T_0+1|T_0}^{+}=e_{T_0+1}-\rho
_{1}e_{T_0}+o_{p}(1)=(\phi _{1}-\rho _{1})e_{T_0-1}+\phi _{2}e_{T_0-2}+o_{p}(1)$,
while the standard
prediction error is $\hat{e}_{T_0+1|T_0}=\phi
_{1}e_{T_0-1}+\phi _{2}e_{T_0-2}+o_p(1)$. Both   have a mean of zero, implying that  misspecifying the
dynamics will not contribute to unconditional bias.  Nonetheless, the \plup\;
variance is  $(1-\rho _{1}^{2})\sigma _{e}^{2}$, which is smaller than the standard prediction error variance of $ \sigma _{e}^{2}$ since $\left\vert \rho _{1}\right\vert \leq 1$. Thus the AR(1) correction  unambiguously   reduces   one-step asymptotic mean squared prediction error. In theory, an AR(p) correction with known parameters  should improve prediction when $e_t$ is an AR(p). But in practice,  sampling uncertainty may offset some gains. Furthermore, when the assumed AR(p) is not the true model,   the mean-squared error is no longer tractable  as shown in \citet{kunitomo-yamamoto:85} even without sampling error. The AR(1) correction is appealing because it is simple to implement, and  precise  mean-squared error statements can be made when there is no sampling uncertainty as stated in Lemma \ref{lem:lemma1}.

Next, consider cases when $h>1$. The standard prediction $\hat{y}_{T_0+h|T_0}=X_{T_0+h}^{\prime }\hat{%
\beta}$ has error  $\hat{e}_{T_0+h|T_0}=y_{T_0+h}-\hat{y}_{T_0+h|T_0}
=e_{T_0+h}+o_{p}(1)$, where $E[e_{T_0+h}]=0$ and $\var(e_{T_0+h})=\gamma
_{0}$.  There are two ways to implement \plup.  The first to  use the AR(1) model for $e_t$ to iteratively  predict $e_{T_0+h}$.    {\em Iterated \plup} (or \plupi),   defined as
\begin{equation*}
\hat{y}_{T_0+h|T_0}^{+I}=X_{T_0+h}^{\prime }\hat{\beta}+\hat{\rho}_{1}^{h}\hat{e}%
_{T_0},
\end{equation*}%
  has error  $\hat{e}_{T_0+h|T_0}^{+I}=e_{T_0+h}-\rho
_{1}^{h}e_{T_0}+o_{p}(1)$. As  shown in the Appendix,
 \[e_{T_0+h}-\rho _{1}^{h}e_{T_0}\dsim (0, \gamma _{0}[1+\rho _{1}^{2h}-2\rho
_{1}^{h}\rho _{h}]).\]
The second approach is to directly predict $e_{T_0+h}$ using information up to $T_0$.   Let $\hat{\rho}_{h}=\hat{\gamma}_{h}/\hat{\gamma}_{0}$ be the $h^{th}$
order sample autocorrelation coefficient of $\{\hat{e}_{t}:t=1,\ldots ,T_0\}$. Direct \plup\; (or \plupd) is defined as
\begin{equation*}
\hat{y}_{T_0+h|T_0}^{+d}=X_{T_0+h}^{\prime }\hat{\beta}+\hat{\rho}_{h}\hat{e}_{T_0}.
\end{equation*}%
 \plupd\; has   error  $\hat{e}_{T_0+h|T_0}^{+d}=e_{T_0+h}-\rho _{h}e_{T_0}+o_{p}\left( 1\right) $, where
\[
e_{T_0+h}-\rho _{h}e_{T_0}\dsim (0, \gamma
_{0}\left( 1-\rho _{h}^{2}\right)) .\]

\begin{lemma}
\label{lem:lemma2} ($h\ge 1)$ Under the same assumptions as in Lemma \ref{lem:lemma1},
the asymptotic MSE of  \plupd\; is always smaller than or equal
to that of \plupi\; and that of the standard predictor for all $h\geq 1$.
\end{lemma}
The proof is  given in the Appendix.
The standard predictor, \plupi\; and \plupd\; are all asymptotically unbiased provided that $E[e_t]=0$.  But
%The Lemma indicates that the asymptotic MSE of \plupd\; can be no larger than  that of \plupi\;  or the standard predictor  at all $h$ and for all data generating processes that satisfy Assumption A1. Precisely,
 \begin{eqnarray*}
\var(e_{T_0+h}-\rho _{h}e_{T_0})-\var\left( e_{T_0+h}-\rho _{1}^{h}e_{T_0}\right)
&=&-\gamma _{0}\left( \rho _{h}-\rho _{1}^{h}\right) ^{2}\leq 0
\end{eqnarray*}%
with equality when the AR(1) model is correctly specified.  Furthermore,
\[\var(e_{T_0+h}-\rho _{h}e_{T_0})-\var(e_{T_0+h})= -\rho_h^2 \gamma_0 \le 0\]
 with equality when $\rho_h=0$.
 Hence,  absent sampling uncertainty, the asymptotic MSE of \plupd\; using  $e_{T_0}$ to improve the standard prediction  can be no larger than \plupi\; which uses the same information for correction, or the standard predictor which ignores $e_{T_0}$ for  any $h\ge 1$.
 This is not to say that a richer dynamic model  would not produce further improvements. What's noteworthy is that  even  a simple  correction will  reduce the prediction  MSE  at any $h$.  We can also expect  the \plupd\; gains  to  be largest at $h=1$ and diminish with $h$ because  the long horizon forecast of a covariance stationary process is the unconditional mean. Though precise statements can be made for \plupd,  we can only say that
% a comparison between \plupi\; and the standard predictor  when $h>1$ requires restricting the class of data generating processes further. We can only say  that
the asymptotic MSE of  \plupi\; is smaller or equal than that of the standard predictor if $1+\rho _{1}^{2h}-2\rho _{1}^{h}\rho _{h}\leq 1$ (see the Appendix for a proof).

Once a prediction is made, we can construct prediction intervals.     We will be studying \plup\;based inference under normality in the context of causal inference. As we will see  in Section 5, while the standard prediction  is unconditionally unbiased, it  is conditional biased and  conditional inference will, in general, have the wrong coverage.

\subsection{Simulations for Linear Predictions}

 This subsection  evaluates the  unconditional and conditional prediction bias, MSE, and coverage  with and without \plup\;correction in a single equation setting  where by unconditional inference, we mean that $e_{i,T_0}$ is random in repeated sampling, and
by conditional inference, we mean that $e_{i,T_0}$ is treated as fixed with respect to  some  conditioning  information, as would be the case in practice.

In each of the 5000 replications, we  first simulate  $K=2$
regressors and  $y_t=X_t'\beta+e_t$  where for $T=1,\ldots, 200$, $e_t=\phi_{1} e_{t-1}+\phi_{2}  e_{t-2}+v_{t}$.   With $v_t\dsim N(0,.05)$,  the $R^2$ of the regression is about 2/3.  In Case 1,  $e_{t}$ is an AR(1)  with $\phi_1=.8$, and in Case 2, $e_t$ is  an AR(2)  with  $(\phi_1,\phi_2)=(1.3,-.4)$.
 Table \ref{tbl:table1} reports four sets of errors in  predicting $y_{T_0+h}$.  The column
labeled 'best' is the infeasible prediction when $\beta ,\phi_1,\phi_2$, and $%
e_{T_0} $ are known.  The column labeled 'noadj' is also infeasible but unlike 'best', it does not take into account information in $e_{T_0}$.  The column labeled 'ols' is the standard prediction $
\hat y_{T_0+h}$ using the  least squares estimate $\hat\beta$.  The  columns  \plupi\;and \plupd\;are iterative and direct \plup\; respectively.  Both are based on a simple AR(1) correction, ie.  even when the true DGP is AR(2).  Note that they are identical when $h=1$.

The top panel of Table \ref{tbl:table1} reports the unconditional
bias and MSE for horizons $h=1,2,5,10$.
The average over all 10 horizons  is reported in the row labeled 'avg'. Since $E[e_{T_0}]=0$,
the unconditional prediction bias is close to zero. However, the
unconditional prediction MSE is much smaller with  \plup\;corrections.   In the AR(1) case, the MSE for the standard (OLS) prediction is  0.14 at $h=1$, but the \plup\;corrections reduce the MSE to  0.05. In the AR(2) case, the OLS prediction has an MSE of  0.43 while the \plup\;corrections reduce the MSE to  0.06. The MSE improvements are smaller when $h>1$, but still non-trivial.

The middle panel of Table \ref{tbl:table1}  shows  conditional prediction errors when  $(e_{T_0-1},e_{T_0})$ are fixed to $(0.5, 1)$.
All  predictions are conditionally biased, but  the \plup\; biases are significantly smaller. When $h=1$ and the errors are AR(1), the standard prediction has a conditional bias of 0.79 while the \plup\;corrections reduce it to 0.01. When the errors are an AR(2) process but an AR(1) correction is implemented, the conditional OLS bias at $h=1$ is reduced from 1.09 to 0.18. Correspondingly, the MSE is reduced from 1.26 to 0.09.  Note that the biases are largest when $h=1$ because the predictability of a stationary ergodic process decreases with the forecast horizon. The improvements in MSE at $h=1$ translate into improved average predictions over 10 periods.  Without the corrections, the average prediction in the AR(2) case has a bias of 0.61 and an MSE of 0.56. The  AR(1) \plup\; direct correction reduces bias to 0.17 and MSE to  0.22.
%Results for $T=200$ reported in the Supplementary Appendix are even more encouraging.

\section{Imputation of Counterfactual Outcomes}

Imputation concerns prediction of values that are never observed.
The problem  is widely studied in a static setup, but  there are few results for  a dynamic setting. \citet[Ch. 11]{little-rubin:19} consider an AR(1) model where $y_1,y_3,\ldots, y_{T-1}$ are observed but not $y_2,y_4,\ldots, y_T$.  The adjustments, shown to require  an implicit regression of $y_t$ on $y_{t-1}$ and $ y_{t+1}$,  can be seen as smoothed estimates of a suitably defined Kalman filter.  \citet{chow-lin:71} consider missing values occurring  between two releases of low frequency data  and show that the best prediction  involves a correction term that has a \blup\;  form.   \citet{ng-scanlan:24} consider factor-based imputation of weekly  missing  values of a scalar series occurring throughout the sample.

Causal inference concerns imputation of missing  potential outcomes that tend to occur at the end of the sample. The problem is typically studied for an iid setting  when it is natural to assume that the errors are uncorrelated\footnote{\citet{causalimpact:15}  consider state space estimation of the counterfactual outcomes in the presence of trends, but  serial correlation in  idiosyncratic shocks and/or the factors are not allowed. \citet{carvalho-masini-medeiros:18,masini/medeiros-jasa:21,masini-medeiros:22} consider causal inference in a high-dimensional setting when the data are persistent and possibly non-stationary.}.
As suggested in Section 2,  correlation in the residuals cannot be ruled out.  We will consider the imputation problem from the perspective of optimal prediction, with the goal of using  the insights of \blup\;  to  improve the imputation of $Y_{i,T_0+h}(0)$. We assume that  $e_{it}=Y_{it}(0)-m_{it}$ are  strong mixing processes and rule out non-stationary data. In addition, we impose the following high level assumption:
\paragraph{Assumptions A2:} For $h\ge 1$, $\hat{m}_{i,T_0+h}-m_{i,T_0+h}=o_{p}\left( 1\right) $
and  $T_{0}^{-1}\sum_{t=1}^{T_{0}}\left( \hat{m}_{it}-m_{it}\right)
^{2}=o_{p}\left( 1\right) $.

Assumption A2 is verified in \citet{cwz-jasa:21} for   estimators  including synthetic control, matrix completion, factor-based methods. Given an asymptotically unbiased $\hat m_{i,T_0+h}$ satisfying Assumptions A1 and A2,  the estimated  treatment effect
 \begin{eqnarray*}
\hat \delta_{i,T_0+h} &=&m_{i,T_0+h}+\delta_{i,T_0+h}+e_{i,T_0+h}-\hat m_{i,T_0+h}
\end{eqnarray*}
has   error
\begin{eqnarray}
\hat \delta_{i,T_0+h}-\delta_{i,T_0+h} &=& (m_{i,T_0+h}-\hat m_{i,T_0+h})+e_{i,T_0+h}. \label{eq:imputation_error}
%&=& \hat e_{1,T_0+h}
\end{eqnarray}
This error has two components: an in-sample estimation uncertainty component that  depends on the estimator but  vanishes as $T_0 \rightarrow \infty$, and an  out-of-sample prediction component that depends on the choice of $m_{it}$ and  the information $\mathcal H$ used in the imputation.

In order to extend Goldberger's  \blup\; from a complete data setting to a potential outcomes setting,
define the $n\times 1$  vector of (observed) control  outcomes $\mathcal Y(0)$   by
\[ \underbrace{\mathcal Y(0)}_{n\times 1}= \begin{pmatrix*}[l] Y_{1,1:T_0} \\ \vdots \\ Y_{N,1:T_0} \\ Y_{N_1+1,T_0+1:T} \\ \vdots \\   Y_{N,T_0+1:T} \end{pmatrix*} \equiv\begin{pmatrix*}[l] \mathcal Y^{pre}(0)_{(NT_0\times 1)} \\ \\ \mathcal Y^{post}(0)_{ (N-N_1)T_1 \times 1} \end{pmatrix*}
\]
where  $n=(NT -N_1T_1)$.  Note that $\mathcal Y(0)$ includes not only the pre-intervention outcomes on all units $\mathcal Y^{pre}(0)\equiv (Y_{1,1:T_0} \ldots Y_{N,1:T_0})'$ but also the post-intervention outcomes on the control units $\mathcal Y^{post}(0)\equiv (Y_{N_1+1,T_0+1:T}\ldots Y_{N,T_0+1:T})'$.

%{\color{blue} Serena, I used italics to define the vectors $Y^{pre}$ and $Y^{post}$, but if you prefer not to use this notation, please change it back. I also think we should define the notation we use. So, I tried to added it here, but then I thought this is obvious...so, I am not inserting it. I did insert the sentence that defines the pre and post notation we use...
%
%Here and throughout, we use the subscript $1:T_0$ to indicate that we collect all the time series observations ranging from $t=1,\ldots,T_0$ into the corresponding vector. Similarly, we use the notation $1:N$ to refer to the cross sectional observations $i=1,\ldots,N$.
%
%}

Let $\mathcal M$ be the pseudo-conditional mean for $\mathcal Y(0)$  and  $\mathcal E$ be the corresponding errors. The matrices $\mathcal M$ and $\mathcal E$ contain typical elements $m_{it}$ and $e_{it}$, respectively. With this notation,
\begin{eqnarray*}
 \mathcal Y(0)&=&\mathcal M +\mathcal E, \quad\quad \mathcal E=\begin{pmatrix} \mathcal E^{pre} \\  \mathcal E^{post} \end{pmatrix} \dsim (0,\Gamma)\\
\Gamma&=& E[\mathcal E \mathcal E'] = \begin{pmatrix}
\Gamma_{pre,pre} & \Gamma_{pre,post} \\
\Gamma_{post,pre} & \Gamma_{post,post}
\end{pmatrix}.
\end{eqnarray*}
The $n\times n$ matrix $\Gamma$  depends on $\Sigma=E[e_te_t']$, which is the $N\times N$ covariance matrix of  $e_t=(e_{1:N_1,t}',e_{N_1+1:N,t}')'$. It also depends   on $\Omega_i=E[e_ie_i']$, which is the time series covariance  of unit $i$, and its  dimension  can be  $T\times T$ or $T_0\times T_0$ depending on whether $i$ is treated.

Consider the linear case   $\mathcal M =\mathcal X\beta$, where $\mathcal X$ contains observed predictors. Consider obtaining  \blup\; given information on $\mathcal Y(0)$ and $\mathcal X$. Re-doing Goldberger's problem gives the following:
\begin{proposition}
\label{prop:prop1}
 Assume that treatment assignment is known, $\mathcal Y(0)=\mathcal M +\mathcal E$, where $\mathcal M =\mathcal X\beta$ where $\beta$ is constant across $i$ and $t$.  Let $\Gamma$ be the $n\times n$ variance-covariance of $\mathcal E$, where $n=NT-N_1T_1$.
  The \blup\; of the counterfactual outcome for unit $i\in[1,N_1]$ is
\[ \mathcal Y_{i,T_0+h}^+(0)=X_{i,T_0+h}'\beta_{GLS}+\omega_{ih}^\prime\Gamma^{-1} \mathcal E_{GLS}\]
where $\beta_{GLS}$ is the vector of infeasible GLS estimates and $\mathcal E_{GLS}$ are the corresponding residuals,  $\omega_{ih}=E[\mathcal E e_{i,T_0+h}]$ is  $n\times 1$ vector of covariances between the unexplained errors in the vector of observed control outcomes and unit $i$'s counterfactual outcome not explained by the model at $T_{0}+h$.
\end{proposition}

Proposition \ref{prop:prop1} provides the individual level best linear unbiased prediction in a treatment effects setting. Two cases are of  special interest.

\paragraph{Case 1: serial correlation only:}  If   $E[e_{\ell t}e_{js}]=0$ for $\ell\ne j$ and for all $t,s$,  then  for $i\in[1,N_1]$,
 \begin{equation}\omega_{ih}'\Gamma^{-1} \mathcal E_{GLS}= \theta_i'e_{GLS,i,1:T_0} \
\label{eq:case1}
\end{equation}
where
$\theta_i =(E[e_{i,1:T_0} e_{i,1:T_0}'])^{-1} E[e_{i,1:T_0}e_{i,1:T_0+h}]$ is the $T_0\times 1$ vector of coefficients from projecting  $e_{i,T_0+h} $ on $(e_{i,1},\ldots, e_{i,T_0})$.

The result in (\ref{eq:case1}) follows from the fact that when there is  no cross-section dependence in the errors, then the  correction for unit $i$ only depends on $\Omega_i$,  the $T_0\times T_0$  autocovariance structure of $e_i$. For instance, if $N_1=1$ and $i=1$ is the treated unit,
 $$\omega_{ih}=\begin{pmatrix} E[e_{i,1:T_0}e_{i,T_0+h}] \\ 0_{N_0T\times 1} \end{pmatrix},$$
where $N_0=N-N_1$, and the prediction simplifies to $ m_{i,T_0+h}+\rho^h_{i}  e_{GLS,i,T_0}$ if $e_{it}$ is assumed to be an  AR(1), where $\rho_i$ is the first order autocorrelation coefficient of $e_{it}$. This coincides with Goldberger's correction reviewed in Section 3.

\paragraph{Case 2: cross-section correlation only:}  If $E[e_{\ell t}e_{js}]=0$ for $t\ne s$ and for all $\ell,j$, then for $i \in [1,N_1]$,
\begin{equation}
\label{eq:case2}
\omega_{ih}'\Gamma^{-1} \mathcal E_{GLS}=\sum_{j=1}^{N-N_1} \theta_{i,N_1+j} e_{GLS,N_1+j,T_0+h}.
\end{equation}
where  $\theta_{i,N_1+j}$ are the slope coefficients from projecting $e_{it}$ on $e_{N_1+1,t},\ldots,e_{N,t}$ using $t=1,\ldots,T_0$.

 \blup\; corrections with cross-section dependence have been derived in a static variance components setting but not in our set up. The result in (\ref{eq:case2}), which is new, is based on two features that follow from no serial correlation. First,  for any treated unit $i\in [1,N_1]$,
$$\omega_{ih}= \begin{pmatrix} E(\mathcal E^{pre}e_{i,T_0+h}) \\ E(\mathcal E^{post}e_{i,T_0+h})\end{pmatrix} = \begin{pmatrix} 0_{NT_0\times 1} \\ E(\mathcal E^{post}e_{i,T_0+h})\end{pmatrix}.$$ Second, the covariance matrix of errors $\Gamma$ has a block diagonal structure
\[ \Gamma=\begin{pmatrix} \Sigma \otimes I_{T_0} & 0 \\ 0 & \Sigma_{00} \otimes I_{T_1}\end{pmatrix}, \quad \Sigma =\begin{pmatrix} \Sigma_{11} & \Sigma_{10} \\ \Sigma_{10}' & \Sigma_{00} \end{pmatrix},\]
 where  $\Sigma_{11}=E(e_{1:N_1,t}e_{1:N_1,t}')$, $\Sigma_{00}=E(e_{N_1+1:N,t}e_{N_1+1:N,t}')$, and $\Sigma_{10}=E(e_{1:N_1,t}e_{N_1+1:N,t}')$. Let $[\Sigma_{10}]_{i,.}$ be the $i$-th row of the matrix $\Sigma_{10}$.
Then, for any $i\in [1,N_1]$ and $h>0$,  the non-zero entries of the $n\times 1$ vector $\omega_{ih}$  are given by
\[ E(\mathcal E^{post}e_{i,T_0+h})= [\Sigma_{01}]_{i.}'\otimes J_h\]
%\[\omega_{ih}=E[\mathcal E e_{i,T_0+h}]=\begin{pmatrix} E[\mathcal E^{pre}\; e_{i,T_0+h}] \\ E[\mathcal E^{post} e_{i,T_0+h}]\end{pmatrix}=\begin{pmatrix} \underbrace{0}_{NT_0\times 1} \\ \underbrace{[\Sigma_{10}]_{i,.}' \otimes J_{h}}_{(N-N_1)T_1\times 1} \end{pmatrix}\]
where   $J_h$ is the $h$-th column of the identity matrix of dimension $T_1$.  As a consequence of the two features, the \blup\; correction  is
\[ \omega_{ih}'\Gamma^{-1} \mathcal E_{GLS}= ([\Sigma_{10}]_{i,.}\Sigma_{00}^{-1} \otimes J_{h}') \mathcal E^{post}_{GLS}. \]
It is a linear combination of $\mathcal E_{GLS}^{post}$, the GLS errors of the control units in the post-treatment sample, with weights given by  the $N-N_1$ vector   $\theta_i'=(\theta_{i,N_1+1},\ldots, \theta_{i,N})=\Sigma_{00}^{-1}[\Sigma_{10}]_{i,.}'$, where     $\theta_{i,N_1+j}$ is the population coefficient associated with $e_{N_1+j,t}$ in the regression of $e_{it}$ on $e_{N_1+1:N,t}=(e_{N_1+1,t},\ldots,e_{N,t})'$.
Given the definition of $J_h$, we can  write the \blup\;correction for unit $i$ in period $T_0+h$ as
$ \omega_{ih}'\Gamma^{-1} \mathcal E_{GLS}= \theta_i'e_{GLS,N_1+1:N,T_0+h}$, as given in the Proposition.
The $\theta_i$ vector can be sparse such as in  factor models when  the idiosyncratic errors can only be weakly dependent in the sense that if $E(e_{it}e_{jt})=\tau_{ij,t}$, $|\tau_{ij,t}|\le |\bar \tau_{ij}|$ for some $\bar \tau_{ij}$ for all $t$, and $\sum_{j=1}^N |\bar \tau_{ij}|\le M\le \infty$ for all $i$.

\subsection{From \textsc{blup} to \textsc{pup}}

In Section 3, we take as a starting point that  the first order improvement of \blup\;comes from controlling for the predictability in  $e_{it}$. While  \blup\; is developed for linear predictions, linearity is not necessary to obtain an improved predictor. Our {\em practical unbiased predictor} (henceforth,  \pup) can be used with any choice of $\mathcal M$ that can be consistently estimated so that the  residuals $\hat{\mathcal E}=\mathcal Y (0)-\hat{\mathcal M}$ can be used to improve prediction.
  In practice, implementation still requires parametric assumptions on $\Omega$ and $\Sigma$. For a serially correlated process $e_{it}$ satisfying mixing conditions,  we have the following

\begin{lemma}
\label{lem:lemma3}
Under Assumptions A1 and A2, a standard imputation has error $ \hat{\delta}_{i,T_{0}+1}-\delta _{i,T_{0}+1}=e_{i,T_{0}+1}+o_{p}(1)$, whose variance is $\sigma^2_{e,i}=\gamma_{0,i}$.
A \pup\; correction has error
 $\hat{\delta}_{i,T_{0}+1}^{+}-\delta _{i,T_{0}+1}=e_{i,T_{0}+1}-\rho
_{i,1}e_{i,T_{0}}+o_{p}(1)$, whose variance is $(\sigma^+_{e,i,1})^2=\gamma_{0,i}(1-\rho^2_{i,1})$, where $\rho_{i,1}$ is the first-order autocorrelation of $e_{it}$. Since $|\rho_{i,1}| \le 1$, $(\sigma^+_{e,i,1})^2 \le \sigma^2_{e,i}$.

\end{lemma}

 Lemma \ref{lem:lemma3} follows immediately from Lemmas  \ref{lem:lemma1} and \ref{lem:lemma2} but the results are presented in a treatment effect setting where the $o_p(1)$ term vanishes with $T_0$. For $h=1$, the MSE of a \pup\;imputation will always be smaller than that of a standard imputation. We focus on $h=1$ since the gain in MSE should be largest, and furthermore, it is also the case when the direct and iterative corrections coincide. In theory, a direct \pup\; using  $\rho_{i,h}$  has better properties than an iterative \pup\;  using  $\rho_{i,1}^h$ when $h>1$. But in simulations when  sampling uncertainty is present,   the two behave similarly and both dominate the standard predictor.

It is also possible to entertain both time and cross-section dependence in $e_{it}$. For example,
\begin{equation}
\label{eq:plup-csts}
\hat Y_{i,T_0+h}^{+}(0)=\hat Y_{i,T_0+h}(0) + \sum_{s=0}^{p_i}\hat\rho_{is}\hat e_{i,T_0-s}+  \sum_{j \ne i}^N  \sum_{s=- p_j}^{p_j}\hat  \theta_{ij,s}  \hat e_{j,T_0+h-s}.
\end{equation}
The first  correction captures the time  series information from  $\hat e_{i}$'s own history, while the second correction  captures cross-section dependence using estimates of the current and past  idiosyncratic errors of the control units.

Though optimal prediction of counterfactual outcomes has not been studied, \pup\;like corrections have recently been considered.
\citet{cwz-jasa:21}  consider time dependence in $e_{1t}$ and suggest adding to the standard imputed value an AR(p) estimate of the residuals, but the idea is not flushed out. \citet{fmm-jasa:22} consider a factor model with observables $W_{it}$ and  assume contemporaneously correlated  idiosyncratic errors  $e_{it}=\theta_i' e_{-i,t}+ v_{it}$, where $e_{-i,t}$ is  a $N-1$ vector that excludes $e_{it}$.  This correlation can be controlled  by $\hat e_{jt}$ from estimation of  the factor model. To circumvent overfitting  the augmented  prediction model
$ Y_{it}(0) =\gamma_i' W_{it}+\lambda_i'F_t+ \theta_1'\hat e_{-i,t}+v_{it}$  when $N_0$ is large,  \textsc{lasso} is used to select which of the  $\hat e_{j,t}$ to keep.     Our analysis  provides a framework for thinking about these \pup\;like modifications.

\subsection{Simulations:  $\hat \delta_{it}$ vs $\hat\delta_{it}^+$}
In Table \ref{tbl:table1}, we saw in a linear prediction  setting  that  \plup\; yields  significant reductions in bias and  mean-squared error.
Since imputation is a form of prediction,  we can  anticipate  improvements  from using \pup\; in imputation settings. To illustrate this, we  generate $Y_{it}(0)$ using a factor model:
\begin{eqnarray*}
Y_{it}(0)&=&c_i+\Lambda_i' F_t+e_{it}, \quad t=1,\ldots, T\\
Y_{1t}(1)&=& Y_{1t}(0)+\delta_{1t}, \quad\quad \delta_{1t}= 0.1 \quad t=T_0+1,\ldots ,T.
\end{eqnarray*}
We  assume that unit 1 is treated with $\delta_{1t}=0.1$ and $e_{it}=\phi_i e_{i,t-1}+v_{it}$, $v_{it}\dsim N(0,.25)$, $\phi_1=0.6$, $\phi_i=0$ for $i>1$. We set $c_i=0$ for all $i$,
  $r=2$, $F_{1t}=.8F_{1t-1}+e^F_{1t}$ and $F_{2t}=.5 F_{2,t-1}+e^F_{2t}$, $\Lambda_{ik}\dsim N(0,1) $, $e^F_{1t}\dsim N(0,.5)$, $e^F_{2t}\dsim N(0,.3)$.  The `best' prediction is $Y_{1,T_0+h}=\Lambda_1'F_{T_0+h}+ \phi_1^h e_{1,T_0}$ and the standard prediction is based on the principal components $\hat \Lambda_i'\hat F_{T_0+h}$ of the demeaned data for the control group, which will be denoted \textsc{pca}.

Table  \ref{tbl:table2} reports the  error in imputing $\hat \delta_{1,T_0+h}$ for $(T,N)=(50,20)$. Because $Y_{1t}(1)$ is observed,  $\hat\delta_{1t}-\delta_{1t}=Y_{1t}(1)-\hat Y_{1t}(0)$ is  due entirely to  $\hat Y_{1t}(0)$.
The top panel considers  $e_{1t}=0.6 e_{1t-1}+v_{1t}$. 
We see that though \textsc{pca} is unconditionally unbiased, its  MSEs are larger compared to the  \pup\; ones.  Furthermore,  the
conditional biases upon fixing $e_{1,T_0}=1.0$ are  smaller with \pup,  as is MSE.  For mutually correlated errors when $e_{1t}=0.5 e_{2t}+v_{1t}$, the conditional case randomly draws  $e_{2,T_0+1:T_1}$ once and keeps the vector fixed in the replications.  Again, the two imputations based on \pup\;  reduce MSE unconditionally and conditionally. Unlike in the time series case, the improvements can occur at any horizon $h$. Though direct \pup\; has slightly better  population properties than iterative \pup\;, they  have rather similar properties in simulations. To simplify notation, we use the term \pup\; in the next subsection, with the understanding that either direct or iterated correlation can be used.

\section{Inference Based on Imputed Counterfactual Outcomes}
 Let $F\left( x\right) =P\left( e_{it}\leq x\right) $ be the marginal distribution function of $%
e_{it}$,   and    $F^+\left(
x\right) $ be the marginal distribution of an adjusted \pup\; error. Lemma 3 suggests that
\begin{eqnarray*}
\hat{\delta}_{i,T_{0}+1}-\delta _{i,T_{0}+1}&\overset{d}{\longrightarrow }&
e_{i,T_0+1}\dsim F\\
\hat{\delta}_{i,T_{0}+1}^{+}-\delta _{i,T_{0}+1}&\overset{d}{
\longrightarrow }&e_{i,T_0+1}-\rho _{i,1}e_{i,T_0}\dsim F^+.
\end{eqnarray*}
%The foregoing analysis suggests that
%\begin{eqnarray*}
%\hat{e}_{i,T_0+1|T_0} &=&y_{T_0+1}-\hat{y}_{i,T_0+1|T_0}\overset{d}{\longrightarrow }%
%e_{T_0+1}\sim F \\
%\hat{e}_{i,T_0+1|T_0}^{+} &=&y_{T_0+1}-\hat{y}_{i,T_0+1|T_0}^{+}\overset{d}{%
%\longrightarrow }e_{i,T_0+1}-\rho _{1}e_{i,T_0}\sim F^+.
%\end{eqnarray*}%
Though both $F$ and $F^+$ are
centered at zero,   $F$ is more dispersed than  $F^+$.
There is surprisingly little work on inference based on a feasible  \blup\; even in a single equation setting presumably because  further  assumptions on $F$ are needed.   We will assume that  $\left\{ e_{it}\right\} $ is a Gaussian process with autocovariance at lag $j$ of $\gamma_{j,i}$, and let $z_{1-\alpha }$ be such that $\Phi \left( z_{1-\alpha }\right)
=1-\alpha$, where $\Phi$ is the cdf of the standard normal distribution.   The Gaussian assumption is only made to illustrate the issues created by omitting predictability. Other distributions can be used in its place so long as $e_{it}$ satisfies our mixing assumption.

We consider  intervals  of the form $\hat\delta\pm \hat\sigma_\delta z_{1-\alpha/2}$ where $\hat\delta$ is either   the standard predictor, or a \pup\; predictor with variance defined in Lemma \ref{lem:lemma3}. These intervals, denoted PI$_{ih}$, and PI$^{+}_{ih}$, can be based on asymptotic theory   or  resampling methods as in  \cite{li-jasa:20} and \citet{cattaneo-jasa:21}, among others.
The intervals will be used for both unconditional and conditional inference. By unconditional inference, we mean that $e_{i,T_0}$ is random in repeated sampling.
By conditional inference, we mean that $e_{i,T_0}$ is treated as fixed with respect to  some  conditioning  information, as would be the case in practice.
 \citet{phillips-joe:79} notes that while unconditional inference is useful for evaluating econometric methods, conditional inference has a role in applications.

We begin with unconditional inference. We say that a prediction interval is \textit{unconditionally} asymptotically valid
if it contains $\delta _{i,T_{0}+h}$ with probability $1-\alpha $ as $%
T_{0}\rightarrow \infty $.
We will focus on pointwise results for unit $i$ where $
e_{i,T_0+h}\equiv Y_{i,T_0+h}( 0) -m_{i,T_0+h}$. It is easy to show that under the assumptions of  Lemma \ref{lem:lemma3} and assuming that $\left\{ e_{it}\right\} $ is  Gaussian, all three intervals are asymptotically valid unconditionally. Consider for instance $\text{\textsc{PI}}_{ih}$, an interval for the standard  prediction error for unit $i$. Since
$\hat{\delta}_{i,T_{0}+h}-\delta _{i,T_{0}+h}=e_{i,T_{0}+h}+o_{p}(1)$
and $e_{i,T_{0}+h}\dsim N\left( 0,\gamma _{0,i}\right) $,  where $\gamma_{0,i}\equiv\sigma^2_{e,i}$, we have
\begin{eqnarray*}
P\left( \delta _{i,T_{0}+h}\in \text{\textsc{PI}}_{ih}\right)  &=&P\left(
\hat{\delta}_{i,T_{0}+h}-\hat{\sigma}_{e,i} z_{1-\alpha /2}\leq \delta
_{i,T_{0}+h}\leq \hat{\delta}_{i,T_{0}+h}+\hat{\sigma}_{e,i} z_{1-\alpha /2}\right)
\\
&=&P\left( -z_{1-\alpha /2}\leq \hat{\sigma}_{e,i}^{-1}(\hat{\delta}%
_{i,T_{0}+h}-\delta _{i,T_{0}+h})\leq z_{1-\alpha /2}\right)  \\
&=&P\left( -z_{1-\alpha /2}\leq \sigma_{e,i} ^{-1}e_{i,T_{0}+h}\leq z_{1-\alpha
/2}\right) +o\left( 1\right)  \\
&=&\Phi \left( z_{1-\alpha /2}\right) -\Phi \left( -z_{1-\alpha /2}\right)
+o\left( 1\right)  \\
&=&1-\alpha +o\left( 1\right).
\end{eqnarray*}%
In the above,  the third equality follows because $\hat{\sigma}_{e,i}^{-1}(\hat{\delta}%
_{i,T_{0}+h}-\delta _{i,T_{0}+h})=\sigma_{e,i} ^{-1}e_{i,T_{0}+h}+o_{p}(1)$ and
the fourth equality uses the Gaussianity assumption on $e_{it}$. The argument for the \pup\;prediction intervals is similar and thus
all three intervals are (unconditionally) asymptotically valid.
In the absence of sampling uncertainty, PI$^{+}$ is asymptotically narrower than PI  for all $h$ whether or not $e_{it}$ is truly an AR(1) process.

\subsection{Conditional Inference}

\bigskip

While  the three intervals provide correct unconditional coverage, will they all have correct conditional coverage? In particular, will they cover $\delta_{i,T_0+h}$ with the nominal coverage probability of $1-\alpha$, conditionally on some information set $\mathcal H$? To answer this question,   consider again the AR(1) case when $e_{it}=\phi_i e_{i,t-1}+v_{it}$, $v_{it}\dsim N(0,\sigma^2_{v,i})$, and $\mathcal H$ is an information set containing $e_{i,T_0}$.
Under the assumption of normality,  $e_{i,T_0+1} \dsim N(\phi_i e_{i,T_0},\sigma^2_{v,i})$ conditionally on $e_{i,T_0}$ and hence
$\frac{e_{i,T_0+1}-\phi_i e_{i,T_0}}{\sigma_{v,i}}\dsim N(0,1)$. But as $  \frac{e_{i,T_0+1}}{\sigma_{e,i}}  \dsim  N(\phi_i \frac{ e_{i,T_0}}{\sigma_{e,i}},\frac{\sigma^2_{v,i}} {\sigma^2_{e,i}}) \ne N(0,1)$, the standard prediction will not usually have the correct coverage  unless $\phi_i$ or $e_{i,T_0}$ are zero. Indeed,
 \textsc{PI}$_{i1}$ will not have the correct conditional
coverage probability even asymptotically because $P\left( \delta _{i,T_{0}+1}\in \text{\textsc{PI}}_{i1}\vert e_{i,T_0}\right)$ is asymptotically equal to
\begin{equation*}
\Phi \left( -\frac{\phi _{i}}{\sigma_{v,i}}{e}_{i,T_0}+\frac{\sigma_{e,i}}{\sigma _{v,i}}z_{1-\alpha /2}\right) -\Phi \left( -\frac{\phi _{i}}{%
\sigma _{v,i}}{e}_{i,T_0}-\frac{\sigma_{e,i}}{\sigma _{v,i}}z_{1-\alpha /2}\right).
\end{equation*}
When $\phi _{i}\neq 0$, the first term  will not return the normal cdf at level $1-\alpha/2$ unless ${e}_{i,T_0}$ is zero.

The problem of distorted inference extends to a multi-period ahead conditional inference.
  For any $h>1$, we see that
\begin{equation*}
e_{i,T_{0}+h}=\phi _{i}^{h}e_{i,T_{0}}+u_{i,T_{0}+h}|e_{i,T_0}\dsim
N\left( \phi _{i}^{h} e_{i,T_{0}},\omega_{h,i}^{2}\right) ,
\end{equation*} where $u_{i,T_{0}+h}\dsim (0,\omega^2_{h,i})$,  $\omega^2_{h,i}=\gamma _{0,i}\left( 1-\phi _{i}^{2h}\right)$,\footnote{This follows because $u_{i,T_{0}+h}=v_{i,T_{0}+h}+\ldots +\phi _{i}^{h-1}v_{i,T_{0}}$. Thus,
 $\omega^2_{h,i}=\sigma _{v,i}^{2}\sum_{j=0}^{h-1} \phi_i^{2h} =\sigma _{v,i}^{2}\frac{1-\phi _{i}^{2h}}{%
1-\phi _{i}^{2}}\equiv \gamma _{0,i}\left( 1-\phi _{i}^{2h}\right)$,
$\gamma _{0,i} =\sigma _{v,i}^{2}\left(
1-\phi _{i}^{2}\right) ^{-1}$.} with $\gamma _{0,i}\equiv\sigma^2_{e,i} =\sigma _{v,i}^{2}\left(
1-\phi _{i}^{2}\right) ^{-1}$. The problem arises because the standard prediction error is not centered at zero  when we condition on $e_{i,T_0}$. Thus for any $h\ge 1$, $P(\delta _{i,T_{0}+h}\in \text{\textsc{PI}}_{ih}|e_{i,T_0})$ is asymptotically equal to
\begin{equation*}
%P^*\bigg(-1.96\sigma_e\le e_{i,T_0+h} \le 1.96\sigma_e\bigg)
\Phi \left(- \frac{\phi_i^h}{\omega_{h,i}}e_{i,T_0}+z_{1-\alpha /2}\frac{\sigma_{e,i}}{\omega_{h,i}}\right)
-\Phi \left(- \frac{\phi_i^h}{\omega_{h,i}}e_{i,T_0}-z_{1-\alpha /2}\frac{\sigma_{e,i}}{\omega_{h,i}}\right)
%%P^*\bigg(\frac{-1.96\sigma_e}{\omega_h}-\phi_1^h \frac{\bar e_{i,T_0}}{\omega_h} \le
%%\frac{e_{i,T_0+h}-\phi_1^h \bar e_{i,T_0}}{\omega_h}
%%\le \frac{1.96\sigma_e}{\omega_h}-\phi_1^h\frac{\bar e_{T_0}}{\omega_h}\bigg)
\end{equation*}
where $\omega^2_{h,i}$ is the h-period forecast error variance defined above. For fixed $h$, conditional inference is distorted unless $e_{i,T_0}= 0$, though  the distortion decreases with $h$ because $\phi_i^h$ tends to zero.

To illustrate the extent of conditional bias, consider two models for $e_{it}$: one when  $e_{it}= \phi_i e_{i,t-1}+v_{it}$ is an AR(1), and one when  $e_{it}=v_{it}+\theta_i v_{i,t-1}$ is an MA(1).  In both cases, the conditional forecast is biased.   Analytically evaluating actual coverage for any  $i$ for  a nominal $95\%$ interval, we have
\begin{center}

Standard Prediction with  $(e_{1,T_0},\sigma_{v,1})=(-2.0,0.5)$

\begin{tabular}{r|rr|rr}  \hline
  h &  coverage & bias & coverage & bias \\ \hline
    &   \multicolumn{2}{c|}{AR(1): $\phi_1=0.8$ } & \multicolumn{2}{c}{MA(1): $\theta_1=0.8$ } \\ \hline
  1 &   0.84 &  -2.26 &   0.58 &  -1.77 \\
  2 &   0.87 &  -1.41 &   0.95 &  -0.00 \\
  3 &   0.90 &  -1.01 &   0.95 &  -0.00 \\
  4 &   0.92 &  -0.76 &   0.95 &  -0.00 \\
  5 &   0.93 &  -0.59 &   0.95 &  -0.00 \\
\hline
\end{tabular}
\end{center}
\bigskip

The coverage probability  is 0.95 at $\phi_1=\theta_1=0$ for all values of $h$ because there is no conditional bias. When $\phi_1\ne 0$ in the AR(1) case, bias  decreases with $\sigma^2_{v,1}$ and $h$. When  $\theta_1\ne 0$ in the MA(1) case, the bias is limited to $h=1$ because the process has a memory of one period. Coverage is distorted by bias, as suggested by theory.

In contrast,  \pup\; has error $\hat{\delta}^{+}_{i,T_{0}+h}-\delta _{i,T_{0}+h}=v_{i,T_{0}+h}+o_{p}(1)$. It is conditionally centered at zero
with conditional variance $\sigma^2_{v,i}$. Let $\hat {\sigma}%
_{i,h}^{+}$ denote a consistent estimator of this variance (which will depend on whether  an iterative or a direct estimator is used).
Under normality, the  \pup\; prediction interval
$\text{PI}_{ih}^{+}=\hat{\delta}_{i,T_{0}+h}^{+}\pm \hat{\sigma}%
_{i,h}^{+}z_{1-\alpha /2}$ has the correct coverage asymptotically.
At $h=1$ when  iterative and direct \pup\; coincide,   $P\left( \delta _{i,T_{0}+1}\in \text{\textsc{PI}}_{i1}^{+}|e_{i,T_0}\right)$ is asymptotically
%$\hat{\delta}_{i,T_{0}+1}^{+}-\delta _{i,T_{0}+1}=e_{i,T_{0}+1}-\phi
%_{1}e_{i,T_{0}}+o_{p}(1)=v_{i,T_{0}+1}+o_{p}(1).$
\begin{equation*}
%P^{\ast }\left(-z_{1-\alpha /2}\leq \frac{\hat{\delta}_{i,T_{0}+1}^{+}-\delta _{i,T_{0}+1}}{ \hat{\sigma}_{i,1}^{+I}}\leq z_{1-\alpha
%/2}\right)  \\
%&\approx &P^{\ast }\left( -\frac{\hat{\sigma}_{i,1}^{+}}{\sigma _{v,i}}%
%z_{1-\alpha /2}\leq \frac{e_{i,T_{0}+1}-\phi _{i}e_{i,T_{0}}}{\sigma _{v,i}}\leq
%\frac{\hat{\sigma}_{i,1}^{+}}{\sigma _{v,i}}z_{1-\alpha /2}\right)  \\
\Phi \left( \frac{\hat{\sigma}_{i,1}^{+}}{\sigma _{v,i}}z_{1-\alpha
/2}\right) -\Phi \left( -\frac{\hat{\sigma}_{i,1}^{+}}{\sigma _{v,i}}z_{1-\alpha
/2}\right)=1-\alpha +o_{p}\left( 1\right) ,
\end{equation*}%
where the last equality uses the fact that $\hat{\sigma}_{i,1}^{+}\pconv \sigma _{v,i}$.  %This result also holds for direct \plup\; since $PI_{i1}^{+d}=PI_{i1}^{+I}$ when $h=1$.

Analogous to  conditional bias due to time dependence, a similar bias occurs when the errors are cross-sectionally correlated. In particular, suppose that $e_{it}$ is serially uncorrelated for all $i$, but  $E[e_{it}e_{jt}]\ne 0$ for at least one $j\ne i$. If $i=1 $ is the only treated unit, the model of interest is   $Y_{1t}(0)=m_{1t}+e_{1t}$ with $e_{1t}=\theta_1'e_{2:N,t}+v_{1t}$.
%Here, $v_{1t}\sim N(0,\sigma^2_{v,1})$, $e_{2:N,t}'=\left(e_{2t},\ldots,e_{Nt}\right)$, and $\theta_1=\Sigma_{00}^{-1}\Sigma_{10}'$, where $\Sigma_{00}=E \left(e_{2:N,t}e_{2:N,t}'\right)$ and $\Sigma_{10}=E \left(e_{1t}e_{2:N,t}'\right)$.
Consider two treatment effects estimators, one based on the standard prediction of $Y_{1t}(0)$ and another based on \pup\;. The standard prediction yields $\hat{\delta}_{1,T_0+1}-\delta_{1,T_0+1}=e_{1,T_0+1}+o_{p}(1)$ which has asymptotic mean zero and asymptotic variance $\sigma^2_{e,1}\equiv E(e^2_{1t})$. Instead, the \pup\;prediction yields a prediction error  $\hat{\delta}^{+}_{1,T_0+1}-\delta_{1,T_0+1}=e_{1,T_0+1}-\theta_1'e_{2:N,T_0+1}+o_{p}(1)$, whose asymptotic variance is $\sigma^2_{v,1}=\sigma^2_{e,1}-\Sigma_{01}'\Sigma_{00}^{-1}\Sigma_{01}<\sigma^2_{e,1}$. With cross-sectionally correlated errors, the standard prediction is conditionally biased because  $e_{1,T_0+1}\vert e_{2:N,T_0+1}\dsim N(\theta_1'e_{2:N,T_0+1},\sigma^2_{v,1})$  is not centered at zero.
Conditional coverage under normality is asymptotically determined by
%\begin{gather*}
% P\bigg(-z_{1-\alpha /2} \le \frac{\hat\delta_{1,T_0+1}-\delta_{1,T_0+1}}{\hat\sigma_{e,1}} \le z_{1-\alpha /2} |e_{2:N,T_0+1}\bigg)\\
%= P \bigg( -\frac{\theta_1'}{\sigma_{v,1}}e_{2:N,T_0+1}-z_{1-\alpha /2} \frac{\sigma_{e,1}}{\sigma_{v,1}} \le v_{1,T_0+1} \le -\frac{\theta_1'}{\sigma_{v,1}}e_{2:N,T_0+1}+z_{1-\alpha /2} \frac{\sigma_{11}}{\sigma_{v,1}}\bigg)\\
%=\Phi\bigg(-\frac{\theta_1'}{\sigma_{v,1}} e_{2:N,T_0+1}+z_{1-\alpha /2}\frac{\sigma_{e,1}}{\sigma_{v,1}}\bigg)-
%\Phi\bigg(-\frac{\theta_1'}{\sigma_{v,1}} e_{2:N,T_0+1}-z_{1-\alpha /2}\frac{\sigma_{e,1}}{\sigma_{v,1}}\bigg).
%\end{gather*}
$$\Phi\bigg(-\frac{\theta_1'}{\sigma_{v,1}} e_{2:N,T_0+1}+z_{1-\alpha /2}\frac{\sigma_{e,1}}{\sigma_{v,1}}\bigg)-
\Phi\bigg(-\frac{\theta_1'}{\sigma_{v,1}} e_{2:N,T_0+1}-z_{1-\alpha /2}\frac{\sigma_{e,1}}{\sigma_{v,1}}\bigg).$$

The conditional bias arising from $\theta_1\ne 0$ and $\sigma_{e,1}\ne \sigma_{v,1}$ will distort inference. Notably,
 $-\frac{\theta_1'}{\sigma_{v,1}}$ here plays the role of $-\frac{\phi_1}{\sigma_{v,1}}$ in the AR(1) setting. But in contrast to the case of serial correlation, the size distortion does not diminish with $h$.  The \pup\; prediction is more efficient because
 $e_{1t}|e_{2:N,t}\dsim (\theta_1'e_{2:N,t},\sigma^2_{v,1})$ which has a smaller  variance.

To illustrate, suppose that $N_1=1$ and $e_{1t}=\theta_1 e_{2t}+v_{1t}$, where we assume that $\sigma_{12}$ is the only non-zero cross sectional covariance. In this case, the relevant parameters are $[\Sigma_{00}]_{11}=\sigma^2_{e,2}$, and  $[\Sigma_{01}]_{11}=\cov(e_{1t},e_{2t})\equiv \sigma_{12}$. Unlike in the time series case when we only condition on $e_{1,T_0}$,  we now condition on $e_{2,T_0+h}$ for each $h$. In the following example, we  draw  $ e_{2,T_0+h}$ once from the normal distribution using the \textsc{rndn} function in \textsc{matlab} with seed 1234, and $\cov(e_{1t},e_{2t})$ from the uniform distribution using the \textsc{rand} function with seed  57.

\newpage
% The program checkcdf.m is in subfolder forc
\begin{center}
Standard Prediction when $e_{1,T_0+h}=\theta_1 e_{2,T_0+h}+v_{1,T_0+h}$, $e_{2,T_0+h}$ fixed\\ $\Sigma=\diag(\sigma^2_{e,1}, \Sigma_{00})=\diag(0.5, 0.841)$.

 \begin{tabular}{rr|rr|rr}  \hline
    &  & \multicolumn{2}{c|}{$\Sigma_{01}=0.613$ } & \multicolumn{2}{c}{$\Sigma_{01}=-0.613$ } \\ \hline
  h &  $\bar e_{2,T_0+h}$ &coverage & bias & coverage & bias \\ \hline
 1  &    0.68 &  0.63 &  -1.64 &   0.89 &  -0.70 \\
  2 &   -0.83 &  0.81 &   1.07 &   0.86 &   0.85 \\
  3 &   -0.92 &  0.93 &  -0.38 &   0.84 &   0.95 \\
  4 &    0.09 &  0.95 &   0.17 &   0.95 &  -0.09 \\
  5 &    0.86 &  0.94 &   0.32 &   0.86 &  -0.89 \\
\hline
\end{tabular}
\end{center}

 The parameterizations result in $\theta_1=\text{sgn} \times 0.72$ where the sign depends on whether $\Sigma_{01}\equiv \sigma_{12}$ is positive or  negative. The sign   affects the magnitude of the bias which in turn affects the extent of  size distortion. However,  unlike in the time series case when predictability falls with $h$ by nature of stationarity, the cross-section correlation does not decrease with $h$. As a consequence, the effect on coverage can vary significantly across horizons.

\subsection{Simulations: PI vs PI$^+$}

Tables \ref{tbl:table1} and \ref{tbl:table2} above showed that \pup\; corrections reduce bias and mean-squared prediction error significantly. But do these improvements  lead to more accurate inference? We first return to Table \ref{tbl:table1}  when point prediction is based on the simple linear model is $y_t=X_t'\beta+e_t$. The corresponding results for coverage are reported  in Table \ref{tbl:table1}(b). Note first that there is no gain in unconditional coverage regardless of the error structure because $E[e_{it}]=0$ by assumption. Hence we focus  on  conditional coverage.  In the AR(1) case, coverage is  40\% at $h=1$ without \pup\;correction, but is at the desired level of 95\% with correction. \pup\; coverage in the AR(2) case is 91\%, which is less accurate, but still  better than the OLS coverage of 0.69.

Evaluation of coverage based on  $\delta_{it}$ is more involved because the sampling error depends on the estimator. Furthermore, given estimates of $\hat\delta_{i,T_0+h}$ for some $i \in [1,N_1]$, any of the following  hypothesis can be considered.
\[
\begin{array}{lllll}
H_0^A&: \delta_{i,T_0+h}&= \delta_{i,T_0+h}^0 \\
H_0^B&:  \delta_{i,T_0+h}&=0  & \forall h\\
 H_{0}^C&: \Delta_{i,T_0+1:T}&= \Delta_{i,T_0+1:T}^0 \quad &\Delta_{i,T_0+1:T}&=\frac{1}{T_1}\sum_{h=1}^{T_1} \delta_{i,T_0+h}\\
 H_{0}^D&: \Delta_{1:N_1,T_0+h}&=\Delta_{1:N_1,T_0+h}^0, \quad  &\Delta_{1:N_1,T_0+h}&=\frac{1}{N_1}\sum_{i=1}^{N_1} \delta_{i,T_0+h}\\
 H_{0}^E&: \Delta_{1:N_1,T_0+1:T}&=\Delta_{1:N_1,T_0+1:T_1}^0, \quad  &\Delta_{1:N_1,T_0+1:T} &=\frac{1}{T_1}\frac{1}{N_1}\sum_{h=1}^{T_1} \sum_{i=1}^{N_1} \delta_{i,T_0+h}
\end{array}
\]
While $A$ and $B$ are pointwise hypotheses, hypotheses C,D,E concern the average treatment effect of the treated, where the average can be taken over time, over units, or  both.  The interpretation of each test depends on $T_1$ and $N_1$. Consider $H_0^C$.
If  $T_1$ is small so that  $\Delta_{i,T_0+1:T}=\frac{1}{T_1}\sum_{t=T_0+1}^T \delta_{it}$ is random,
  we construct a {\em prediction} interval for $\Delta_{i,T_0+1:T}$. When  $T_1$ is large,   $\Delta_{i,T_0+1:T}\pconv \Delta_{i,\infty}=E[\delta_{it}]$   is  non-random.   In this case, we  construct a {\em confidence} interval  for  $\Delta_{1,T_0+1:T}$.  Considerations of $N_1$ are likewise needed for testing   $H_0^D$.\footnote{ For large $T_1$,
$  \frac{1}{T_1}\sum_{h=1}^{T_1}  m_{i,T_0+h}-\hat m_{i,T_0+h}+ \frac{1}{T_1}\sum_{h=1}^T e_{i,T_0+h}+ \delta_{i,T_0+h}-E[\delta_{i,T_0+h}]
$, which equals
$\hat {\Delta}_{i,T_0+1:T}-\Delta_{i,T_0+1:T}+  \delta_{i,T_0+h}-E[\delta_{i,T_0+h}].$
}

To give a flavor of the results,  we only consider $H_0^A$ and $H_0^C$ using the same data generating process in Table \ref{tbl:table2}. We estimate   $(F_t,\Lambda_{i})$    using the  \textsc{tall-wide}  procedure in \citet{baing-jasa:21}. With this estimator,  the asymptotic distribution of  $\hat \delta_{i,T_0+h}$ is determined by the distribution of $e_{i,T_0+h}$. Hence the distribution of $\hat \delta_{i,T_0+h}$  is normal only if $e_{it}$ is Gaussian. However, the treatment effect averaged over $T_1$ periods can be asymptotically normal with a convergence rate of $\min(T_0,T_1)$ if  the CLT  $\frac{1}{\sqrt{T_0}} \sum_{t=1}^{T_0}  F_te_{it}\dconv N(0,\Phi_i)$ holds. In contrast,  the average treatment effect over $N_1$ units can be asymptotically normal with a convergence rate of $\min(N_0,N_1)$ if  a CLT for $\frac{1}{\sqrt{N_0}}\sum_{i=1}^{N_0}\Lambda_i e_{it}\dconv N(0,\Gamma_t)$ holds.  These distinctions  will help understand the coverage results reported in Table \ref{tbl:table2}(b).

Turning first to  the time series case in the top panel of Table \ref{tbl:table2}(b), \pup\; coverage for $\delta_{i,T_0+h}$  improves over PCA, but only for  $h=1,2$,  suggesting that for serially correlated errors, \pup\; is most effective when  $h$ is small, especially when  $T_0$ is small.  For cross-sectionally correlated errors  reported in the bottom panel, improved coverage of \pup\;   can occur at any $h$, but is more systematic when $N_0$ is large.

As explained above,  the limiting distribution of the \citet{baing-jasa:21} estimator of $\hat{\bar\delta_i}$ depends on adequacy of central limit theory, while it is normality of $e_{it}$ that renders  $\hat\delta_{it}$  normally distributed. It is thus not surprising that
 coverage of $\delta_{it}$ does not provide a good guide to the coverage of $\bar \delta_i=\frac{1}{T_1} \sum_{h=1}^{T_1} \delta_{i,T_0+h}$.  At $(T_0,N_0)=(50,20)$, \pup\; conditional coverage of $\bar\delta_i$ is too low, though  no more inaccurate than the standard \textsc{pca} prediction.
Whether the errors are serially or cross-sectionally correlated, coverage of $\bar\delta_i$ is more reliable as $T_0$ increases because we average over $T_1$ variables that become increasingly Gaussian.   Sampling methods  could be useful when $T_0$ is small, see e.g. \citet{li-shen-zhou:23}.

\section{Conclusion}
 \citet{goldberger:62} shows that if the errors of the prediction model are non-spherical, they can be exploited to improve prediction. Motivated by this result, this paper has suggested \pup, a simple way to adjust existing estimates of counterfactual outcomes for dependence in the errors. The adjustment consists of adding a term that exploits the presence of serial or cross sectional correlation in the prediction errors of the model used to obtain the estimated counterfactual outcomes.

 We showed that improved mean-squared errors are possible without knowledge of the true error structure, and simple  corrections often suffice. We also showed that omitting the \pup\; adjustment term when the error is predictable can result in conditional bias, thus leading to prediction intervals that are not conditionally asymptotically valid. In contrast, a prediction interval based on \pup\; is conditionally unbiased, resulting in valid inference both conditionally and unconditionally.

 While improved predictions are possible, it should be pointed out that when serial correlation is a concern, we can focus on  corrections at small $h$  because $\rho^h$ will be small  for stationary mixing processes. Furthermore, dependence in the residuals is  necessary but  not sufficient for improved prediction.  This is because the \pup\; adjustment   depends not just on $\hat \rho_{is}$ or $\hat \rho_{js}$, but also on the values of $\hat e_{i,T_0-s}$ and  $\hat e_{j,T_0+j-s}$ which can take on values  close to zero.
 In the German unification example, the cross-section \pup\; correction does not make much difference. However,  the residuals exhibit serial dependence and the  AR(1) \pup\;correction  changed the imputed growth rate for 1991 from -1.722 to -1.537 and for 1996 from -1.528 to  -1.662.      If take log GDP as the outcome variable instead of GDP growth,  the residuals are still serially correlated with $\hat \rho_1=0.72$. But $(\hat e_{1,T_0},\hat e_{1,T_0-1})=(0.025,-0.0024)$ which  are small relative to  $y_{1,T_0}=9.26$. Hence in this case, the \pup\; corrections did not make appreciable difference. Ultimately, whether the \pup\; corrections are large is an empirical matter. Our goal is simply to draw awareness to the possibility of improvements. A practical first step could be to  use the in-sample residuals $\hat e_{it}$ to construct   an LM test for  no dependence using  the auxiliary regression
\[ \hat e_{it} = X_{it}'\delta_0+\sum_{s=0}^{p_i} \delta_{i,s}\hat e_{i,t-s}+\sum_{j=1}^N \sum_{s=-p_j}^{p_j} \delta_{j,s} \hat e_{j,t-s}.\]
We can entertain \pup\; corrections if  the null hypothesis is rejected, keeping in mind that the significance of the corrections depend not only on the hypothesis to be tested, but also on the estimator used.

\clearpage
\appendix\setcounter{section}{0}

\section{Appendix: Proofs}

{\bf Proof of Lemma \ref{lem:lemma1}. }The proof of part (i) follows from
standard arguments, see e.g., Hayashi (2000), p. 145. We provide a brief
sketch here. Write $\hat{\rho _{1}}=\frac{\hat{\gamma}_{1}}{\hat{\gamma}_{0}}
$, with $\hat{\gamma}_{j}=\frac{1}{T_0}\sum_{t=1}^{T_0}\hat{e}_{t}\hat{e}_{t-j}$%
, for $j=0,1$. Let $\tilde{\gamma}_{j}=\frac{1}{T_0}\sum_{t=1}^{T_0}e_{t}e_{t-j}$
and note that $\tilde{\gamma}_{j}\rightarrow _{p}\gamma _{j}$ under
Assumption A1. Since $\hat{e}_{t}=y_{t}-X_{t}^{\prime }\hat{\beta}%
=e_{t}+X_{t}^{\prime }(\beta -\hat{\beta})$, we can write%
\begin{equation*}
\hat{\gamma}_{j}=\tilde{\gamma}_{j}-T_0^{-1}%
\sum_{t=1}^{T_0}(X_{t-j}e_{t}+X_{t}e_{t-j})^{\prime }(\hat{\beta}-\beta )+(%
\hat{\beta}-\beta )^{\prime }(T_0^{-1}\sum_{t=1}^{T_0}X_{t}X_{t-j}^{\prime
})^{-1}(\hat{\beta}-\beta ).
\end{equation*}%
The last two terms are $o_{p}\left( 1\right) $ under the assumption that $%
\hat{\beta}-\beta =o_{p}\left( 1\right) $ and assuming that $E(X_{t-j}e_{t})$%
, $E(X_{t}e_{t-j}^{\prime })$ and $E\left( X_{t}X_{t-j}^{\prime }\right) $
are all finite for $j=0,1$. It follows that $\hat{\gamma}_{j}=\tilde{\gamma}%
_{j}+o_{p}(1)=\gamma _{j}+o_{p}\left( 1\right) $, which implies the result.
To prove (ii), note that $\hat{e}_{T_0+1|T_0}=y_{T_0+1}-X_{T_0+1}^{\prime }\hat{\beta%
}=e_{T_0+1}-X_{T_0+1}^{\prime }(\hat{\beta}-\beta )=e_{T_0+1}+o_{p}\left( 1\right)
$ under the assumption that $\hat{\beta}-\beta =o_{p}(1)$. It follows
immediately that $E(e_{T_0+1})=0$ and $Var(e_{T_0+1})\equiv \gamma _{0}=\sigma
_{e}^{2}$. For (iii), note that we can write $\hat{e}_{T_0+1|T_0}^{+}=e_{T_0+1}-%
\rho _{1}e_{T_0}+o_{p}(1)$, given (i) and the assumption that $\hat{\beta}%
-\beta =o_{p}(1)$. Since $E\left( e_{t}\right) =0$ for all $t$, we have $%
E(e_{T_0+1}-\rho _{1}e_{T})=0$. In addition, $Var(e_{T_0+1}-\rho
_{1}e_{T})=\gamma _{0}+\rho _{1}^{2}\gamma _{0}-2\rho _{1}\gamma _{1}=\gamma
_{0}-\gamma _{1}^{2}/\gamma _{0}\leq \gamma _{0}$, where the second equality
follows by replacing $\rho _{1}=\gamma _{1}/\gamma _{0}$.

\textbf{Proof of Lemma~2.} The standard OLS prediction error is%
\begin{equation*}
\hat{e}_{T_0+h|T_0}=y_{T_0+h}-\hat{y}_{T_0+h|T_0}=e_{T_0+h}-X_{T_0+h}^{\prime }(\hat{\beta}%
-\beta )=e_{T_0+h}+o_{p}\left( 1\right) \text{.}
\end{equation*}%
The mean of $e_{T_0+h}$ is zero and the variance is $\gamma _{0}$. The $h$%
-step ahead direct \plup\;based on an AR(1) approximation is%
\begin{equation*}
\hat{e}_{T_0+h|T_0}^{+d}=y_{T_0+h}-\hat{y}_{T_0+h|T_0}^{+d}=e_{T_0+h}-\hat{\rho}_{h}%
\hat{e}_{T_0}+o_{p}\left( 1\right) =e_{T_0+h}-\rho _{h}e_{T_0}+o_{p}\left(
1\right) \text{,}
\end{equation*}where $\hat{e}_{T_0}$ is
the OLS residual for observation $T_0$, and $\hat{\rho}_h$ is the $h$-th order sample autocorrelation of $\hat{e}_{t}$.
It follows immediately that the asymptotic MSE of the direct \plup\; is smaller or equal than that of OLS since the variance of $e_{T_0+h}-\rho _{h}e_{T_0}$ is $\gamma_0(1-\rho_h^2)\le \gamma_0$, since $\rho_h^2\le 1$. To compare direct \plup\; with iterated \plup\;, note that the $h$-step ahead iterated \plup\;based on an AR(1) approximation is%
\begin{equation*}
\hat{y}_{T_0+h|T_0}^{+I}=\hat{y}_{T_0+h|T_0}+\hat{\rho}_{1}^{h}\hat{e}_{T_0},
\end{equation*}%
where $\hat{\rho}_{1}$ is defined in (\ref{eq:rho_hat}) and $\hat{e}_{T_0}$ is
the OLS residual for observation $T_0$. The iterated \plup\;error is%
\begin{equation*}
\hat{e}_{T_0+h|T_0}^{+I}=y_{T_0+h}-\hat{y}_{T_0+h|T_0}^{+I}=e_{T_0+h}-\hat{\rho}_{1}^{h}%
\hat{e}_{T_0}+o_{p}\left( 1\right) =e_{T_0+h}-\rho _{1}^{h}e_{T_0}+o_{p}\left(
1\right) \text{.}
\end{equation*}%
Both versions of \plup\;have mean zero but the variance of iterated \plup\;is larger or equal than that of direct \plup\;since we can show that the difference between these two variances is $\gamma_0(\rho_h-\rho_1^h)^2\ge 0$. A comparison between iterated \plup\;and the standard prediction reveals that both prediction errors have mean zero, but the
variance of the standard OLS prediction is $Var\left( e_{T_0+h}\right) =\gamma
_{0}$ whereas that of the iterated \plup\;error is equal to
\begin{eqnarray*}
Var\left( e_{T_0+h}-\rho _{1}^{h}e_{T_0}\right) &=&\gamma _{0}+\rho
_{1}^{2h}\gamma _{0}-2\rho _{1}^{h}\gamma _{h} \\
&=&\gamma _{0}[1+\rho _{1}^{2h}-2\rho _{1}^{h}(\frac{\gamma _{h}}{\gamma _{0}%
})] \\
&=&\gamma _{0}[1+\rho _{1}^{2h}-2\rho _{1}^{h}\rho _{h})],
\end{eqnarray*}%
since $\rho _{h}=\frac{\gamma _{h}}{\gamma _{1}}$, where $\gamma
_{h}=E\left( e_{t+h}e_{t}\right) $. When
$1+\rho _{1}^{2h}-2\rho _{1}^{h}\rho _{h}\leq 1$
we can conclude that the mean square prediction error of \plup\;is smaller
than the mean square prediction error of OLS.
%Thus, the mean is zero and the variance is $\gamma_0+\rho _{h}^2\gamma_0-2\rho_{h}\gamma_{h}$.

% abadie_data/sng_fbi_fd.r
%\begin{figure}
%\hspace*{-1.0in}
%\caption{Effect of German Unification on  GDP Growth}
%\label{fig:germany-growth}
%\begin{center}
%\includegraphics[width=6.0in,height=4.0in]{../abadie_data/sng_fbi_fd_growth.png}
%\end{center}
%\end{figure}

\clearpage

%% run do_table1.m
%% do_table11.tex -- unconditional results
%% do_table12.tex -- conditional results
%% do_table1_ci.tex -- coverage
\vspace*{-.3in}
\begin{table}[hbt!]
\setlength{\tabcolsep}{4pt}
\caption{Prediction Errors
}
\label{tbl:table1}

\vspace*{-.35in}

{\footnotesize
\begin{center}

\begin{eqnarray*}
 y_t&=& X_{1t}\beta_1+X_{2t}\beta_2+e_t,\quad  \beta_1=\beta_2=1, \quad  T=50\\
X_{1t}&=&0.6 X_{1,t-1}+ u_{1t}, \quad u_{1t}\sim N(0,1), \quad\quad
X_{2t}= u_{2t}, \quad u_{2t}\sim N(0,1)\\
e_t&=& \phi_1 e_{t-1}+\phi_2 e_{t-2}+v_t, \quad v_t\sim N(0,0.05)
\end{eqnarray*}
\end{center}
}
{\footnotesize
\begin{center}

Unconditional

\begin{tabular}{r|rrrrr|rrrrrrr}
h &  best &  noadj & ols & plupI & plupD   & best &  noadj & ols &  plupI & plupD  \\ \hline

&  \multicolumn{5}{c}{bias} & \multicolumn{5}{c}{mse} \\\hline 
  \hline 
 & \multicolumn{10}{c}{($\phi_1,\phi_2$)=(0.80,0.00)}\\  \hline 
    1 &-0.00 & -0.01 & -0.00 & -0.00 & -0.00 &  0.05 &  0.14 &  0.14 &  0.05 &  0.05\\
    2 & 0.00 & -0.00 &  0.00 &  0.00 &  0.00 &  0.05 &  0.14 &  0.14 &  0.08 &  0.08\\
    5 & 0.00 &  0.00 &  0.00 &  0.00 &  0.00 &  0.05 &  0.13 &  0.14 &  0.12 &  0.12\\
   10 & 0.00 & -0.00 & -0.00 & -0.00 &  0.00 &  0.05 &  0.14 &  0.14 &  0.14 &  0.14\\
  avg &-0.00 & -0.00 & -0.00 & -0.00 & -0.00 &  0.00 &  0.07 &  0.08 &  0.06 &  0.06\\\hline 
 & \multicolumn{10}{c}{($\phi_1,\phi_2$)=(1.30,-0.40)}\\  \hline 
    1 &-0.00 & -0.01 & -0.00 & -0.00 & -0.00 &  0.05 &  0.43 &  0.43 &  0.06 &  0.06\\
    2 & 0.00 & -0.00 & -0.00 & -0.00 & -0.00 &  0.05 &  0.43 &  0.43 &  0.16 &  0.16\\
    5 & 0.00 & -0.00 &  0.00 & -0.00 &  0.00 &  0.05 &  0.42 &  0.43 &  0.37 &  0.35\\
   10 & 0.00 & -0.01 & -0.00 & -0.00 & -0.00 &  0.05 &  0.43 &  0.44 &  0.48 &  0.44\\
  avg &-0.00 & -0.00 & -0.00 & -0.00 & -0.00 &  0.00 &  0.28 &  0.29 &  0.21 &  0.20\\\hline 

%\end{tabular}

 \multicolumn{10}{c}{Conditional on $e_{T_0}=1.0$ and $e_{T_0-1}=0.5$} \\ \hline

 & \multicolumn{5}{c}{ bias }  & \multicolumn{5}{c}{mse} \\  \hline 
 & \multicolumn{10}{c}{($\phi_1,\phi_2$)=(0.80,0.00)}\\  \hline 
    1 &-0.00 &  0.80 &  0.79 &  0.01 &  0.01 &  0.05 &  0.68 &  0.68 &  0.05 &  0.05\\
    2 & 0.00 &  0.64 &  0.64 &  0.01 &  0.02 &  0.05 &  0.49 &  0.49 &  0.09 &  0.09\\
    5 & 0.00 &  0.33 &  0.33 &  0.01 &  0.02 &  0.05 &  0.23 &  0.23 &  0.13 &  0.13\\
   10 & 0.00 &  0.11 &  0.10 &  0.00 &  0.02 &  0.05 &  0.15 &  0.15 &  0.14 &  0.15\\
  avg &-0.00 &  0.36 &  0.35 &  0.01 &  0.02 &  0.00 &  0.18 &  0.18 &  0.06 &  0.06\\\hline 
 & \multicolumn{10}{c}{($\phi_1,\phi_2$)=(1.30,-0.40)}\\  \hline 
    1 &-0.00 &  1.10 &  1.09 &  0.18 &  0.18 &  0.05 &  1.25 &  1.26 &  0.09 &  0.09\\
    2 & 0.00 &  1.03 &  1.03 &  0.18 &  0.24 &  0.05 &  1.19 &  1.20 &  0.17 &  0.20\\
    5 & 0.00 &  0.62 &  0.62 & -0.04 &  0.20 &  0.05 &  0.72 &  0.73 &  0.34 &  0.38\\
   10 & 0.00 &  0.21 &  0.21 & -0.23 &  0.08 &  0.05 &  0.47 &  0.48 &  0.49 &  0.45\\
  avg &-0.00 &  0.61 &  0.61 & -0.04 &  0.17 &  0.00 &  0.56 &  0.56 &  0.19 &  0.22\\\hline

\end{tabular}
\end{center}
}
5000 replications using Matlab R2019a, seed = rng(1234,'twister')
\end{table}
%& \multicolumn{10}{c}{Coverage} \\ \hline

\begin{table}[hbt!]]
\begin{center}
Table 1(b) Coverage

%\begin{eqnarray*}
% y_t&=& X_{1t}\beta_1+X_{2t}\beta_2+e_t,\quad  \beta_1=\beta_2=1, \quad  T=50\\
%X_{1t}&=&0.6 X_{1,t-1}+ u_{1t}, \quad u_{1t}\sim N(0,1), \quad\quad
%X_{2t}= u_{2t}, \quad u_{2t}\sim N(0,1)\\
%e_t&=& \phi_1 e_{t-1}+\phi_2 e_{t-2}+v_t, \quad v_t\sim N(0,0.05)
%\end{eqnarray*}
%\end{center}
%\begin{center}

T=50

\begin{tabular}{r|rrrrr|rrrrrrr}
h &  best &  noadj & ols & plupI & plupD   & best &  noadj & ols &  plupI & plupD  \\ \hline

 & \multicolumn{5}{c}{Unconditional} & \multicolumn{5}{c}{Conditional}  \\  \hline 
 & \multicolumn{8}{c}{($\phi_1,\phi_2$)=(0.80,0.00)}\\  \hline 
    1 & 0.96 &  0.95 &  0.95 &  0.95 &  0.95 &  0.96 &  0.40 &  0.41 &  0.95 &  0.95\\
    2 & 0.95 &  0.95 &  0.95 &  0.95 &  0.95 &  0.95 &  0.64 &  0.63 &  0.95 &  0.95\\
    5 & 0.96 &  0.95 &  0.95 &  0.95 &  0.95 &  0.96 &  0.88 &  0.88 &  0.95 &  0.94\\
   10 & 0.95 &  0.95 &  0.95 &  0.94 &  0.94 &  0.95 &  0.94 &  0.94 &  0.94 &  0.94\\
    avg & 0.95 &  0.99 &  0.99 &  0.94 &  0.94 &  0.96 &  0.94 &  0.93 &  0.94 &  0.94\\\hline 
 & \multicolumn{8}{c}{($\phi_1,\phi_2$)=(1.30,-0.40)}\\  \hline 
    1 & 0.96 &  0.95 &  0.95 &  0.95 &  0.95 &  0.96 &  0.79 &  0.77 &  0.92 &  0.92\\
    2 & 0.95 &  0.95 &  0.95 &  0.95 &  0.95 &  0.95 &  0.76 &  0.75 &  0.94 &  0.92\\
    5 & 0.96 &  0.95 &  0.95 &  0.95 &  0.95 &  0.96 &  0.87 &  0.87 &  0.96 &  0.93\\
   10 & 0.95 &  0.95 &  0.94 &  0.95 &  0.94 &  0.95 &  0.94 &  0.94 &  0.94 &  0.94\\
    avg & 0.95 &  0.98 &  0.98 &  0.94 &  0.94 &  0.96 &  0.94 &  0.93 &  0.96 &  0.92\\\hline 
\hline

& \multicolumn{10}{c}{T=200}\\  \hline

 & \multicolumn{4}{c}{Unconditional} & \multicolumn{4}{c}{Conditional}  \\  \hline 
 & \multicolumn{8}{c}{($\phi_1,\phi_2$)=(0.80,0.00)}\\  \hline 
    1 & 0.96 &  0.95 &  0.95 &  0.95 &  0.95 &  0.96 &  0.39 &  0.40 &  0.95 &  0.95\\
    2 & 0.95 &  0.94 &  0.94 &  0.94 &  0.94 &  0.96 &  0.63 &  0.63 &  0.94 &  0.93\\
    5 & 0.95 &  0.95 &  0.94 &  0.94 &  0.94 &  0.96 &  0.87 &  0.86 &  0.93 &  0.92\\
   10 & 0.95 &  0.94 &  0.93 &  0.93 &  0.92 &  0.96 &  0.93 &  0.92 &  0.94 &  0.91\\
  avg & 0.94 &  0.99 &  0.98 &  0.92 &  0.92 &  0.95 &  0.93 &  0.91 &  0.91 &  0.89\\\hline 
 & \multicolumn{8}{c}{($\phi_1,\phi_2$)=(1.30,-0.40)}\\  \hline 
    1 & 0.96 &  0.95 &  0.94 &  0.95 &  0.95 &  0.96 &  0.73 &  0.69 &  0.91 &  0.91\\
    2 & 0.95 &  0.94 &  0.94 &  0.94 &  0.94 &  0.96 &  0.72 &  0.70 &  0.93 &  0.90\\
    5 & 0.95 &  0.94 &  0.93 &  0.94 &  0.93 &  0.96 &  0.85 &  0.84 &  0.94 &  0.91\\
   10 & 0.95 &  0.93 &  0.92 &  0.93 &  0.91 &  0.96 &  0.93 &  0.91 &  0.93 &  0.91\\
   avg & 0.94 &  0.98 &  0.97 &  0.94 &  0.91 &  0.95 &  0.92 &  0.90 &  0.94 &  0.89\\\hline 
\hline

\hline
\end{tabular}
\end{center}
\end{table}

\begin{table}[ht]
\caption{Errors in Estimated Treatment Effect :  $T=50,N=20$}
\label{tbl:table2}
\vspace*{-.35in}
{\footnotesize
\begin{center}
 \begin{eqnarray*}
Y_{it}(0)&=&c_i+\Lambda_i'F_t+e_{it}, \quad \delta_1=0.1, 1\\
e_{it}&=& \phi_i e_{it-1}+v_{it}, \quad v_{it}\sim N(0,0.25), \quad \phi_i=0.6, \phi_j=0, j>1\\
F_{1t}&=& 0.8 F_{1,t-1}+e_{1t}^F, \quad e_{1t}^F\sim N(0,0.5), \quad \Lambda_{1i}\sim N(0,1)\\
 F_{2t}&=& 0.5 F_{2,t-1}+e_{2t}^F, \quad e_{2t}^F\sim N(0,.3), \quad \Lambda_{2i}\sim N(0,1). 
\end{eqnarray*}

\begin{tabular}{r|rrrrr|rrrrrr}
h &    best  &  noadj &  pca &  plupI & plupD  & best & noadj & pca & plupI & plupD  \\ \hline

 & \multicolumn{8}{c}{ $e_{1t}=0.6e_{1t-1}+\epsilon_{1t}$}\\ \hline 
  & \multicolumn{5}{c}{Unconditional bias of $\hat\delta_{i,T_0+h}$} & \multicolumn{4}{c}{mse} \\\hline \
    1 & 0.00 & -0.00 &  0.00 &  0.01 &  0.01 &  0.25 &  0.37 &  0.43 &  0.33 &   0.33\\
    2 &-0.01 & -0.01 & -0.00 & -0.00 & -0.00 &  0.33 &  0.37 &  0.45 &  0.42 &   0.42\\
    5 &-0.00 & -0.00 &  0.01 &  0.01 &  0.01 &  0.37 &  0.39 &  0.48 &  0.47 &   0.48\\
   10 &-0.01 & -0.01 & -0.00 & -0.00 & -0.00 &  0.39 &  0.40 &  0.50 &  0.50 &   0.51\\
  avg &-0.00 & -0.00 &  0.00 &  0.00 &  0.00 &  0.12 &  0.13 &  0.18 &  0.17 &   0.17\\\hline 

 & \multicolumn{8}{c}{ $e_{1t}=0.6e_{1t-1}+\epsilon_{1t}$, $e_{T_0}$=1.00}\\ \hline 
 & \multicolumn{5}{c}{Conditional bias of $\hat\delta_{i,T_0+h}$} & \multicolumn{4}{c}{mse}\\ \hline  
    1 & 0.00 &  0.60 &  0.56 &  0.15 &  0.15 &  0.25 &  0.61 &  0.66 &  0.36 &   0.36\\
    2 &-0.01 &  0.35 &  0.32 &  0.12 &  0.14 &  0.33 &  0.45 &  0.53 &  0.45 &   0.47\\
    5 &-0.00 &  0.21 &  0.19 &  0.09 &  0.13 &  0.37 &  0.42 &  0.51 &  0.48 &   0.52\\
   10 &-0.01 &  0.12 &  0.09 &  0.04 &  0.09 &  0.39 &  0.40 &  0.51 &  0.50 &   0.54\\
  avg &-0.00 &  0.15 &  0.12 &  0.04 &  0.09 &  0.12 &  0.14 &  0.19 &  0.17 &   0.19\\\hline

\hline\hline
 & \multicolumn{8}{c}{ $e_{1t}=0.5e_{2t}+\epsilon_{1t}$}\\ \hline 
  & \multicolumn{5}{c}{Unconditonal bias of $\hat\delta_{i,T_0+h}$} & \multicolumn{4}{c}{mse} \\\hline \
    1 & 0.00 & -0.00 &  0.00 &  0.00 &  0.00 &  0.15 &  0.20 &  0.33 &  0.28 &   0.28\\
    2 &-0.01 & -0.01 & -0.00 & -0.01 & -0.01 &  0.15 &  0.21 &  0.35 &  0.29 &   0.29\\
    5 & 0.00 & -0.00 &  0.01 &  0.01 &  0.01 &  0.15 &  0.21 &  0.35 &  0.30 &   0.30\\
   10 &-0.01 & -0.01 & -0.01 &  0.00 &  0.00 &  0.15 &  0.21 &  0.34 &  0.29 &   0.29\\
  avg &-0.00 & -0.00 &  0.00 &  0.01 &  0.01 &  0.01 &  0.02 &  0.05 &  0.04 &   0.04\\\hline 

&\multicolumn{8}{c}{ $e_{1t}=0.5e_{2t}+\epsilon_{1t}$,  $e_{2T_0+1}=0.27539$} \\  \hline
& \multicolumn{5}{c}{Conditional bias of $\hat\delta_{i,T_0+h}$} & \multicolumn{4}{c}{mse}\\ \hline 
    1 & 0.00 &  0.14 &  0.14 &  0.04 &  0.04 &  0.15 &  0.17 &  0.29 &  0.27 &   0.27\\
    2 &-0.01 &  0.49 &  0.49 &  0.12 &  0.12 &  0.15 &  0.38 &  0.52 &  0.35 &   0.35\\
    5 & 0.00 &  0.47 &  0.48 &  0.13 &  0.13 &  0.15 &  0.37 &  0.52 &  0.35 &   0.35\\
   10 &-0.01 & -0.45 & -0.44 & -0.14 & -0.14 &  0.15 &  0.35 &  0.48 &  0.34 &   0.34\\
  avg &-0.00 & -0.12 & -0.12 & -0.06 & -0.06 &  0.01 &  0.03 &  0.06 &  0.05 &   0.05\\\hline

\hline 

\end{tabular}
\end{center}
5000  replications using Matlab 2019a with seed rng(456,'twister').
For cross-section correlation, $e_{2,T_0:1:T}$=(0.6361, -0.6386,    0.7118,   -1.7044,   -1.2992,    1.6402,    0.1395,    0.9348,    0.5051,    0.9692).
}
\end{table}

\clearpage 

\begin{table}
%\label{tbl:table2-coverage}

{\small
\begin{center}

Table 2(b), Coverage

\begin{tabular}{r|rrrrr|rrrrrr}
h &    best  &  noadj &  pca &  plupI & plupD  & best & noadj & pca & plupI & plupD  \\ \hline

&\multicolumn{10}{c}{$e_{1t}=\phi e_{1t-1}+v_{1t}$}\\ \hline

& \multicolumn{5}{c}{($T_0,N_0$)=(50,20):Unconditional} & \multicolumn{4}{c}{Conditional}  \\  \hline 
    1 & 0.95 &  0.94 &  0.92 &  0.93 &  0.93 &  0.96 &  0.87 &  0.86 &  0.92 &  0.92\\
    2 & 0.91 &  0.94 &  0.91 &  0.91 &  0.91 &  0.92 &  0.92 &  0.90 &  0.91 &  0.90\\
    5 & 0.88 &  0.93 &  0.90 &  0.88 &  0.87 &  0.90 &  0.94 &  0.91 &  0.89 &  0.87\\
   10 & 0.89 &  0.94 &  0.90 &  0.84 &  0.83 &  0.91 &  0.94 &  0.91 &  0.85 &  0.84\\
 avg  & 0.98 &  0.85 &  0.75 &  0.77 &  0.75 &  0.98 &  0.83 &  0.73 &  0.76 &  0.72\\\hline 

 & \multicolumn{5}{c}{($T_0,N_0$)=(200,50): Unconditional  } & \multicolumn{4}{c}{Conditional}  \\  \hline 
    1 & 0.95 &  0.95 &  0.94 &  0.95 &  0.95 &  0.96 &  0.90 &  0.89 &  0.95 &  0.95\\
    2 & 0.91 &  0.95 &  0.94 &  0.94 &  0.94 &  0.91 &  0.92 &  0.92 &  0.94 &  0.94\\
    5 & 0.89 &  0.95 &  0.94 &  0.93 &  0.93 &  0.89 &  0.95 &  0.94 &  0.94 &  0.93\\
   10 & 0.89 &  0.95 &  0.94 &  0.94 &  0.93 &  0.90 &  0.95 &  0.94 &  0.94 &  0.93\\
 avg  & 0.99 &  0.93 &  0.92 &  0.92 &  0.92 &  0.99 &  0.92 &  0.90 &  0.92 &  0.91\\\hline

\hline\hline

&\multicolumn{10}{c}{$e_{1t}=\theta_1 e_{2t}+v_{1t}$}\\ \hline

  & \multicolumn{5}{c}{($T_0,N_0$)=(50,20): Unconditional} & \multicolumn{4}{c}{Conditional}  \\  \hline 
    1 & 0.95 &  0.95 &  0.94 &  0.93 &  0.93 &  0.95 &  0.97 &  0.95 &  0.93 &  0.93\\
    2 & 0.95 &  0.95 &  0.93 &  0.93 &  0.93 &  0.95 &  0.86 &  0.87 &  0.90 &  0.90\\
    5 & 0.95 &  0.94 &  0.93 &  0.91 &  0.91 &  0.95 &  0.91 &  0.91 &  0.90 &  0.90\\
   10 & 0.95 &  0.95 &  0.93 &  0.88 &  0.88 &  0.95 &  0.98 &  0.96 &  0.89 &  0.89\\
 avg  & 0.88 &  0.88 &  0.79 &  0.78 &  0.78 &  0.88 &  0.81 &  0.75 &  0.77 &  0.77\\\hline

 & \multicolumn{5}{c}{ ($T_0,N_0$)=(200,50): Unconditional} & \multicolumn{4}{c}{Conditional}  \\  \hline 
    1 & 0.95 &  0.96 &  0.95 &  0.95 &  0.95 &  0.95 &  0.94 &  0.94 &  0.95 &  0.95\\
    2 & 0.95 &  0.95 &  0.94 &  0.94 &  0.94 &  0.95 &  0.94 &  0.93 &  0.94 &  0.94\\
    5 & 0.95 &  0.95 &  0.95 &  0.95 &  0.95 &  0.95 &  0.75 &  0.78 &  0.93 &  0.93\\
   10 & 0.95 &  0.95 &  0.95 &  0.94 &  0.94 &  0.95 &  0.86 &  0.87 &  0.93 &  0.93\\
 avg  & 0.94 &  0.94 &  0.92 &  0.92 &  0.92 &  0.94 &  0.92 &  0.91 &  0.92 &  0.92\\\hline

\hline
\end{tabular}

\end{center}
}
\end{table}

  \clearpage
\baselineskip=12.0pt
\bibliographystyle{econometrica}
\bibliography{../write/blup}

%\includepdf[pages=-]{blup_table_more.pdf}
\end{document}